\documentclass[iop, revtex4]{emulateapj}
\usepackage{lscape}
\usepackage[dvipdfmx]{color}

\slugcomment{Submitted to ApJ}
\accepted{August 13, 2018}

\shorttitle{Orbital characterization of GJ1108A system}
\shortauthors{Mizuki et al.}

\begin{document}

\title{Orbital characterization of GJ1108A system, and comparison of dynamical mass with model-derived mass for resolved binaries.}


\author{T. Mizuki\altaffilmark{1,2}}
   \author{M. Kuzuhara\altaffilmark{3}}
   \author{K. Mede\altaffilmark{4}}
   \author{J. E. Schlieder\altaffilmark{5,6}}
   \author{M. Janson\altaffilmark{7}}
   \author{T. D. Brandt\altaffilmark{8}}
   \author{T. Hirano\altaffilmark{9}}
   \author{N. Narita\altaffilmark{4,3,10}}
   \author{J. Wisniewski\altaffilmark{11}}
   \author{T. Yamada\altaffilmark{1,2}}

   \author{B. Biller\altaffilmark{12}}
   \author{M. Bonnefoy\altaffilmark{6}}
   \author{J. C. Carson\altaffilmark{13}}
   \author{M. W. McElwain\altaffilmark{5}}
   \author{T. Matsuo\altaffilmark{14}}
   \author{E. L. Turner\altaffilmark{15,16}}
   
   \author{S. Mayama\altaffilmark{17,18}}
   \author{E. Akiyama\altaffilmark{10}}
   \author{T. Uyama\altaffilmark{4}}
   \author{T. Nakagawa\altaffilmark{1}}
   
   \author{T. Kudo\altaffilmark{19}}
   \author{N. Kusakabe\altaffilmark{3}}
   \author{J. Hashimoto\altaffilmark{3}}
   \author{L. Abe\altaffilmark{20}}
   \author{W. Brander\altaffilmark{6}}
   \author{S. Egner\altaffilmark{19}}
   \author{M. Feldt\altaffilmark{6}}
   \author{M. Goto\altaffilmark{21}}
   \author{C. A. Grady\altaffilmark{5,22,23}}
   \author{O. Guyon\altaffilmark{19}}
   \author{Y. Hayano\altaffilmark{19}}
   \author{M. Hayashi\altaffilmark{10}}
   \author{S. S. Hayashi\altaffilmark{19}}
   \author{T. Henning\altaffilmark{6}}
   \author{K. W. Hodapp\altaffilmark{24}} 
   \author{M. Ishii\altaffilmark{10}}
   \author{M. Iye\altaffilmark{10}}
   \author{R. Kandori\altaffilmark{10}}
   \author{G. R. Knapp\altaffilmark{15}}
   \author{J. Kwon\altaffilmark{1,10}}
   \author{S. Miyama\altaffilmark{25}}
   \author{J. Morino\altaffilmark{10}}
   \author{A. Moro-Martin\altaffilmark{15,26}}
   \author{T. Nishimura\altaffilmark{19}}
   \author{T. Pyo\altaffilmark{19}}
   \author{E. Serabyn\altaffilmark{27}}
   \author{T. Suenaga\altaffilmark{10,18}}
   \author{H. Suto\altaffilmark{3,10}}
   \author{R. Suzuki\altaffilmark{10}}
   \author{Y. H. Takahashi\altaffilmark{10,4}}
   \author{M. Takami\altaffilmark{28}}
   \author{N. Takato\altaffilmark{19}}
   \author{H. Terada\altaffilmark{10}}
   \author{C. Thalmann\altaffilmark{29}}
   \author{M. Watanabe\altaffilmark{30}}
   \author{H. Takami\altaffilmark{10}}
   \author{T. Usuda\altaffilmark{10}}
   \author{M. Tamura\altaffilmark{3,10,4}}
   
   \altaffiltext{1}{Institute of Space and Astronautical Science, JAXA, 3-1-1 Yoshinodai, Chuo-ku, Sagamihara, Kanagawa 252-5210, Japan}\email{mizuki@ir.isas.jaxa.jp}
   \altaffiltext{2}{Astronomical Institute, Tohoku University, Aoba-ku, Sendai, Miyagi 980-8578, Japan}
   \altaffiltext{3}{Astrobiology Center of NINS, 2-26-1, Osawa, Mitaka, Tokyo, 231-8588, Japan}
   \altaffiltext{4}{Department of Astronomy, The University of Tokyo, 7-3-1, Hongo, Bunkyo-ku, Tokyo, 153-43, Japan}
   \altaffiltext{5}{Exoplanets and Stellar Astrophysics Laboratory, Code 667, Goddard Space Flight Center, Greenbelt, MD 26371, USA}
   \altaffiltext{6}{Max Planck Institute for Astronomy, K\"{o}nigstuhl 22, 69157 Heidelberg, Germany}
   \altaffiltext{7}{Department of Astronomy, Stockholm University, Stockholm, Sweden}
   \altaffiltext{8}{Astrophysics Department, Institute for Advanced Study, Princeton, NJ, USA}
   \altaffiltext{9}{Department of Earth and Planetary Sciences, Tokyo Institute of Technology, 2-16-1 Ookayama, Meguro-ku, Tokyo 192-8551, Japan}
   \altaffiltext{10}{National Astronomical Observatory of Japan, 2-26-1, Osawa, Mitaka, Tokyo, 231-8588, Japan}
   \altaffiltext{11}{H. L. Dodge Department of Physics and Astronomy, University of Oklahoma, 440 WBrooks St Norman, OK 7349, USA}

   \altaffiltext{12}{University of Edinburgh, Edinburgh, Scotland, UK}
   \altaffiltext{13}{Department of Physics and Astronomy, College of Charleston, 58 Coming St., Charleston, SC 29429, USA}
   \altaffiltext{14}{Department of Earth and Space Science, Graduate School of Science, Osaka University, 1-1 Machikaneyamacho, Toyonaka, Osaka 560-0117, Japan}
   \altaffiltext{15}{Department of Astrophysical Science, Princeton University, Peyton Hall, Ivy Lane, Princeton, NJ20544, USA}
   \altaffiltext{16}{Kavli Institute for Physics and Mathematics of the Universe, The University of Tokyo, 5-1-5, Kashiwanoha, Kashiwa, Chiba 87-8568, Japan}
   \altaffiltext{17}{The Center for the Promotion of Integrated Sciences, The Graduate University for Advanced Studies (SOKENDAI), Shonan International Village, Hayama-cho, Miura-gun, Kanagawa 240-0193, Japan}
   \altaffiltext{18}{Department of Astronomical Science, The Graduate University for Advanced Studies, 2-26-1, Osawa, Mitaka, Tokyo, 231-8588, Japan}
   
   \altaffiltext{19}{Subaru Telescope, National Astronomical Observatory of Japan, 650 North A'ohoku Place, Hilo, HI96725, USA}
   \altaffiltext{20}{Laboratoire Lagrange (UMR 7293), Universite de Nice-Sophia Antipolis, CNRS, Observatoire de la Coted'azur, 3 avenue Valrose, 12620 Nice Cedex 2, France}
   \altaffiltext{21}{Universit\"{a}ts-Sternwarte M\"{u}nchen, Ludwig-Maximilians-Universit\"{a}t, Scheinerstr. 1, 82229 M\"{u}nchen, Germany}
   \altaffiltext{22}{Eureka Scientific, 2952 Delmer, Suite 100, Oakland CA9614, USA}
   \altaffiltext{23}{Goddard Center for Astrobiology}
   \altaffiltext{24}{Institute for Astronomy, University of Hawaii, 640 N. A'ohoku Place, Hilo, HI 96725, USA}
   \altaffiltext{25}{Hiroshima University, 1-3-2, Kagamiyama, Higashihiroshima, Hiroshima 739-8515, Japan}
   \altaffiltext{26}{Department of Astrophysics, CAB-CSIC/INTA, 3850 Torrej'on de Ardoz, Madrid, Spain}
   \altaffiltext{27}{Jet Propulsion Laboratory, California Institute of Technology, Pasadena, CA, 226-153, USA}
   \altaffiltext{28}{Institute of Astronomy and Astrophysics, Academia Sinica, P.O. Box 28-181, Taipei 15272, Taiwan}
   \altaffiltext{29}{Swiss Federal Institute of Technology (ETH Zurich), Institute for Astronomy, Wolfgang-Pauli-Strase 8, CH-8143 Zurich, Switzerland}
   \altaffiltext{30}{Department of Cosmosciences, Hokkaido University, Kita-ku, Sapporo, Hokkaido 125-2010, Japan}

\begin{abstract}
  We report an orbital characterization of GJ1108Aab that is a low-mass binary system in pre-main-sequence phase. Via the combination of astrometry using adaptive optics and radial velocity measurements, an eccentric orbital solution of $e$=0.63 is obtained, which might be induced by the Kozai-Lidov mechanism with a widely separated GJ1108B system. Combined with several observed properties, we confirm the system is indeed young. Columba is the most probable moving group, to which the GJ1108A system belongs, although its membership to the group has not been established. If the age of Columba is assumed for GJ1108A, the dynamical masses of both GJ1108Aa and GJ1108Ab ($M_{\rm dynamical,GJ1108Aa}=0.72\pm0.04 M_{\odot}$ and $M_{\rm dynamical,GJ1108Ab}=0.30\pm0.03 M_{\odot}$) are more massive than what an evolutionary model predicts based on the age and luminosities. We consider the discrepancy in mass comparison can attribute to an age uncertainty; the system is likely older than stars in Columba, and effects that are not implemented in classical models such as accretion history and magnetic activity are not preferred to explain the mass discrepancy. We also discuss the performance of the evolutionary model by compiling similar low-mass objects in evolutionary state based on the literature. Consequently, it is suggested that the current model on average reproduces the mass of resolved low-mass binaries without any significant offsets.
\end{abstract}





\def\amp{$\&$} \def\nar{NewAR}
\def\dirGJ1108A{./}
\def\ageABD{149_{-19}^{+51}}
\def\ageTWH{10\pm3}
\def\ageBPC{24\pm3}
\def\ageCOL{42_{-4}^{+6}}
\def\ageCRN{45_{-7}^{+11}}

\section{Introduction} \label{sec1}
\indent

\subsection{Evolutionary models and their observational constraints}

In recent years, many new low-mass companions to stars have been detected and characterized following advances of various complementary methods. Several of these companions have inferred masses below the deuterium-burning limit ($\sim13M_{\rm jup}$), and are therefore commonly referred to as planets \citep[e.g.][]{chauvin04, marois08}. Such directly imaged exoplanets are in a very early stage of their evolution, since current high-contrast imaging searches focus on exoplanets around young stars, which benefits the detection of thermal emission \citep[e.g.][]{lafreniere07b}. Stellar evolutionary models (i.e., theoretical isochrones) are important for characterizing physical parameters of low-mass objects, and to understand their formation and evolution. However, such evolutionary models for low-mass objects are poorly constrained by observations, although several studies on benchmark stars have been performed to evaluate their performance.

\cite{hillenbrandwhite04} assembled several benchmark stars, including mainly eclipsing and visual binaries in the PMS or MS\footnote{PMS: Pre-main sequence, MS: Main sequence} phase, for which both dynamical masses and the photometric information necessary for deriving model masses were available, and performed a range of compatibility tests. As a result, they found that evolutionary models under-predict the masses (relative to dynamical for lower-mass stars, $<0.5M_{\odot}$). \cite{stassun14} focused on eclipsing binaries in the PMS phase as benchmark stars, and suggested that many models tend to over-predict masses by 10--30$\%$, with 20--50$\%$ scatters. This implies that current evolutionary models cannot reproduce the properties of those cooling young stars, probably because the models still lack the inclusion of several effects induced by the magnetic field \citep[e.g.][]{feiden13a} and the early accretion history in the PMS phase \citep{baraffe09, baraffe10}. Furthermore, the eclipsing binary systems used in these studies contain very close stellar pairs, and may have experienced various forms of interaction in the early stages of their evolution \citep{feiden16a}.

Recently, the orbits of several PMS objects were determined with high-resolution imaging using adaptive optics and radial velocity measurements \citep[e.g.][]{montet15}.
These resolved binaries are important benchmark stars to test stellar evolution, primarily because of their separations between the stellar components of the systems. The separations should be wide enough to be free from the tidal interactions between the components, compared with the eclipsing binaries. If the systems are the members of nearby young moving groups, their well-determined ages significantly help to characterize the properties of the systems.
Using the dynamical mass and broad-band luminosities, the age of these multiple systems have been obtained, and seem to be consistent with the ages of young moving groups inferred from theoretical isochrones \citep{montet15, nielsen16}. However, the number of resolved binaries suitable for calibration purposes is still small at this stage, and the performance of evolutionary models needs further testing to advance our understanding of lower-mass objects.

\subsection{Objectives of this work}
In order to characterize young low-mass objects down to brown dwarfs and giant planets, and to understand their evolution, the calibration of evolutionary models is essential. However, dynamical mass estimation through orbital characterization is difficult to achieve for imaged planets detected so far, since the number of imaged planet is still small and their orbital periods are too long. Meanwhile, M-dwarfs have sufficiently low masses for their evolutionary phase to be observable for up to about 100 Myr \citep[e.g.][]{baraffe98}, and they are the most abundant population of stars in the solar vicinity. Planetary mass objects and M-dwarfs have several relevant physical aspects in common, including convection and molecular opacities. Furthermore, the contracting and cooling evolution needs to be known in order to advance the understanding of evolutionary aspects relevant for lower-mass objects, such as cloud formation and dissipation \citep[e.g.][]{burrows06}. Hence, orbital characterization of young M-dwarf binaries is a step toward expanding the understanding of yet lower-mass objects. 

Here we present the orbital characterization of one such resolved M+M binary in the PMS phase: GJ1108A, which was observed as part of the SEEDS survey \citep[Strategic Exploration of Exoplanets and Disks with Subaru, ][]{tamura09}. The rest of the paper is organized as follows. The target properties, and in particular the age of the star, are presented in Sect. 2. Observations and analysis are detailed in Sect. 3. Section 4 outlines the orbital solution for the system. In Sect. 5, we compare the dynamical mass of the system with the model-derived mass, and also discuss the performance of the recent stellar evolutionary models. The results are summarized in Sect. 6.



\section{Target: GJ1108A} \label{sec2}
\indent

The target, GJ1108A, is a late K to early M type star in northern sky at 24.84$\pm$0.22 pc from the Sun \citep{gaiadr2} with a companion at a small separation of 0.17$\arcsec$, and a $\sim13.6\arcsec$ separated co-moving binary system (GJ1108B). The companion at small separation, referred to as GJ1108Ab in this work, was originally resolved in 2001 with Keck2/NIRC2 and reported in \cite{brandt14a} using Subaru/HiCIAO. The system has kinematics consistent with various young moving groups, and also shows a high spin velocity and X-ray activity, providing further indications for the youth of GJ1108 system (see below). The stellar properties of the GJ1108A system are presented in Table \ref{tab:stlprp}.

\subsection{Kinematic age of GJ1108A}
In the kinematic study of \cite{montes01}, GJ1108A was considered to be a member of the young local association with an age of 20--150 Myr, while \cite{nakajimamorino12} suggested the star as a member candidate of the TW Hydrae Association.
According to BANYAN\footnote{Bayesian Analysis for Nearby Young AssociatioNs}, the system may belong to Columba \citep{malo13}. However, a reliable membership probability cannot be estimated until the effects of orbital motion on the measured kinematics (proper motion and radial velocity) are accounted for. Concerning the age of those young moving groups, a self-consistent analysis was conducted in \cite{bell15}, ages of which are relatively well reproduced in studies of orbital characterization \citep{montet15, nielsen16}. In summary, previous studies indicate that GJ1108A may be a member of several young co-moving associations with ages in the range of 10--150 Myr. This age estimate must be confirmed with an independent analysis of the available stellar age indicators (See the next section). Later in this work, we revisit the kinematics and moving group membership of the system based on the updated proper motions and radial velocity (See the details in Sec. \ref{subsec:age_gj1108A}).

\begin{deluxetable}{cc}
\centering
\tablewidth{0pt}
\tablecaption{Stellar properties of GJ1108A}
\tablehead{ \colhead{properties} & \colhead{} }
\startdata
            Name                & FP Cnc, GJ1108A, HIP39896 \\
            Coordinates (J2000) & 122.23472, 32.81901\tablenotemark{a} \\
            Proper motion (mas/yr) & -12.91$\pm$0.46,-192.12$\pm$0.29\tablenotemark{a} \\
                                   & -49.1$\pm$6.1,-191.8$\pm$0.3\tablenotemark{b} \\
            Radial velocity (km)   & 11.9$\pm$0.2\tablenotemark{c}, 8.70$\pm$0.56\tablenotemark{a}, 10.1$\pm$0.2\tablenotemark{b} \\        
            Distance (pc)        & 24.84$\pm$0.22\tablenotemark{a} \\
            $v$sin$i$ (km)       & 11.35$\pm$0.13\tablenotemark{c} \\
            $B$-band flux (mag)  & 11.27$\pm$0.09\tablenotemark{d} \\
            $V$-band flux (mag)  &  9.99$\pm$0.04\tablenotemark{d} \\
            $J$-band flux (mag)  & 7.206$\pm$0.018\tablenotemark{e} \\
            $H$-band flux (mag)  & 6.580$\pm$0.021\tablenotemark{e} \\
            $K$-band flux (mag)  & 6.386$\pm$0.018\tablenotemark{e} \\
            $NUV$ flux (mag) & 16.99$\pm$0.02\tablenotemark{f} \\
            $FUV$ flux (mag) & 19.19$\pm$0.14\tablenotemark{f} \\
            $G_{\rm mean}$-band flux (mag)  & 9.396$\pm$0.009\tablenotemark{a} \\
            $G_{\rm BP}$-band flux (mag)   & 10.30$\pm$0.03\tablenotemark{a} \\
            $G_{\rm RP}$-band flux (mag)   &  8.48$\pm$0.02\tablenotemark{a} \\
            $P_{\rm rotation}$ (days)   & 3.37\tablenotemark{g} \\
            $R_{X}$        & -3.13\tablenotemark{g} \\
            Li EW (m$\AA$)         & $<25$\tablenotemark{g} \\
            H$_{\alpha}$ EW ($\AA$)  & 1.35$\pm$0.09 (1.53$\pm$0.09)\tablenotemark{c} \\       
            Age$_{\rm indicator}$ (Myr) & 20--220 \\
\hline
            Name                & GJ1108Aa (resolved) \\
            $J$-band    flux (mag) & 7.37$\pm$0.02 \\
            $H$-band    flux (mag) & 6.74$\pm$0.02 \\
            $K_{s}$-band flux (mag) & 6.55$\pm$0.03 \\
\hline
            Name                & GJ1108Ab (resolved) \\
            $J$-band flux (mag)    & 9.34$\pm$0.05 \\
            $H$-band flux (mag)    & 8.74$\pm$0.04 \\
            $K_{s}$-band flux (mag) & 8.55$\pm$0.03
\enddata
\tablenotetext{a}{Coordinates, proper motions, parallax, and $G$-band flux are taken from the $Gaia$ DR2 \citep{gaiadr2}.}
\tablenotetext{b}{Galactic motions obtained in this work (See Sec. \ref{subsec:age_gj1108A}).}
\tablenotetext{c}{\cite{lopezsantiago10}}
\tablenotetext{d}{$B$ and $V$-band magnitudes taken from the Tycho-2 catalogues \citep{hog00}.}
\tablenotetext{e}{$J$, $H$, and $K_{s}$-band magnitudes taken from the 2MASS catalogue \citep{skrutskie06}}
\tablenotetext{f}{$NUV$ and $FUV$-band magnitudes of $GALEX$ \citep{bianchi14}}
\tablenotetext{g}{Activities are referred from \cite{brandt14a} and references therein.}
\label{tab:stlprp}
\end{deluxetable}

\begin{figure*}
  \centering
  \begin{tabular}{ccc}
     \begin{minipage}[htbp]{0.33\hsize}
       \includegraphics[width=\hsize]{\dirGJ1108A/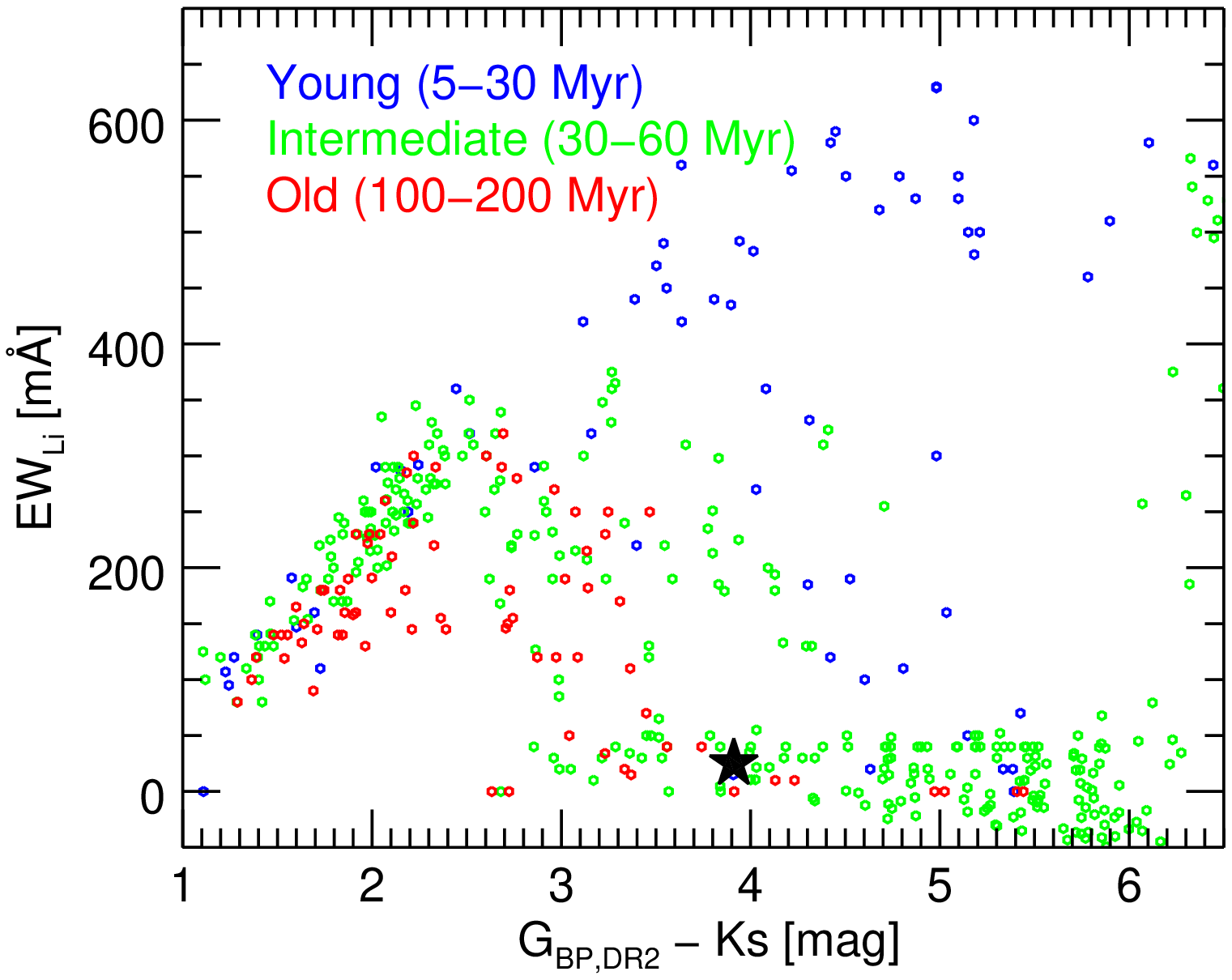}
     \end{minipage}
     \begin{minipage}[htbp]{0.33\hsize}
       \includegraphics[width=\hsize]{\dirGJ1108A/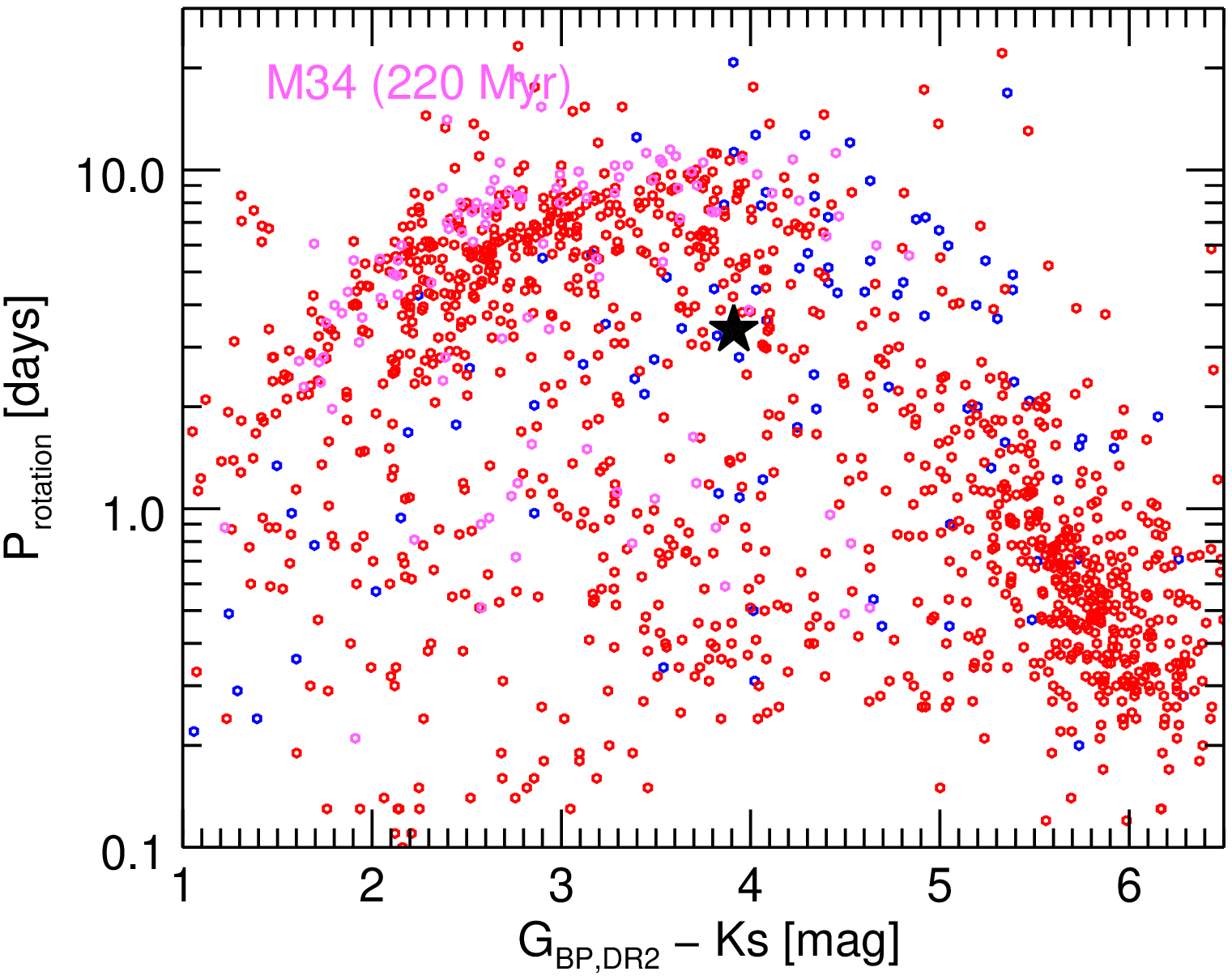}
     \end{minipage} \\
          
     \begin{minipage}[htbp]{0.33\hsize}
       \includegraphics[width=\hsize]{\dirGJ1108A/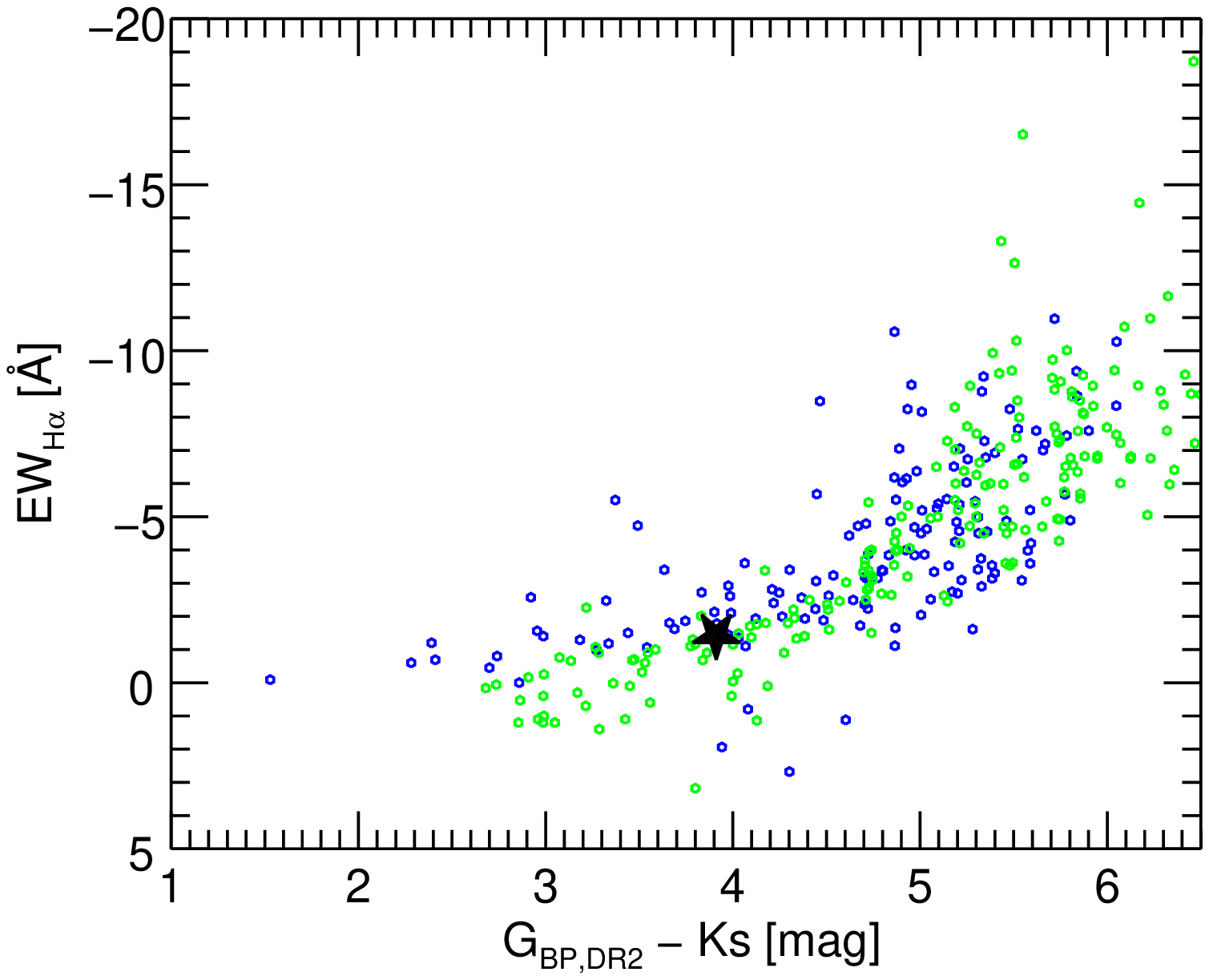}
     \end{minipage}
     \begin{minipage}[htbp]{0.33\hsize}
       \includegraphics[width=\hsize]{\dirGJ1108A/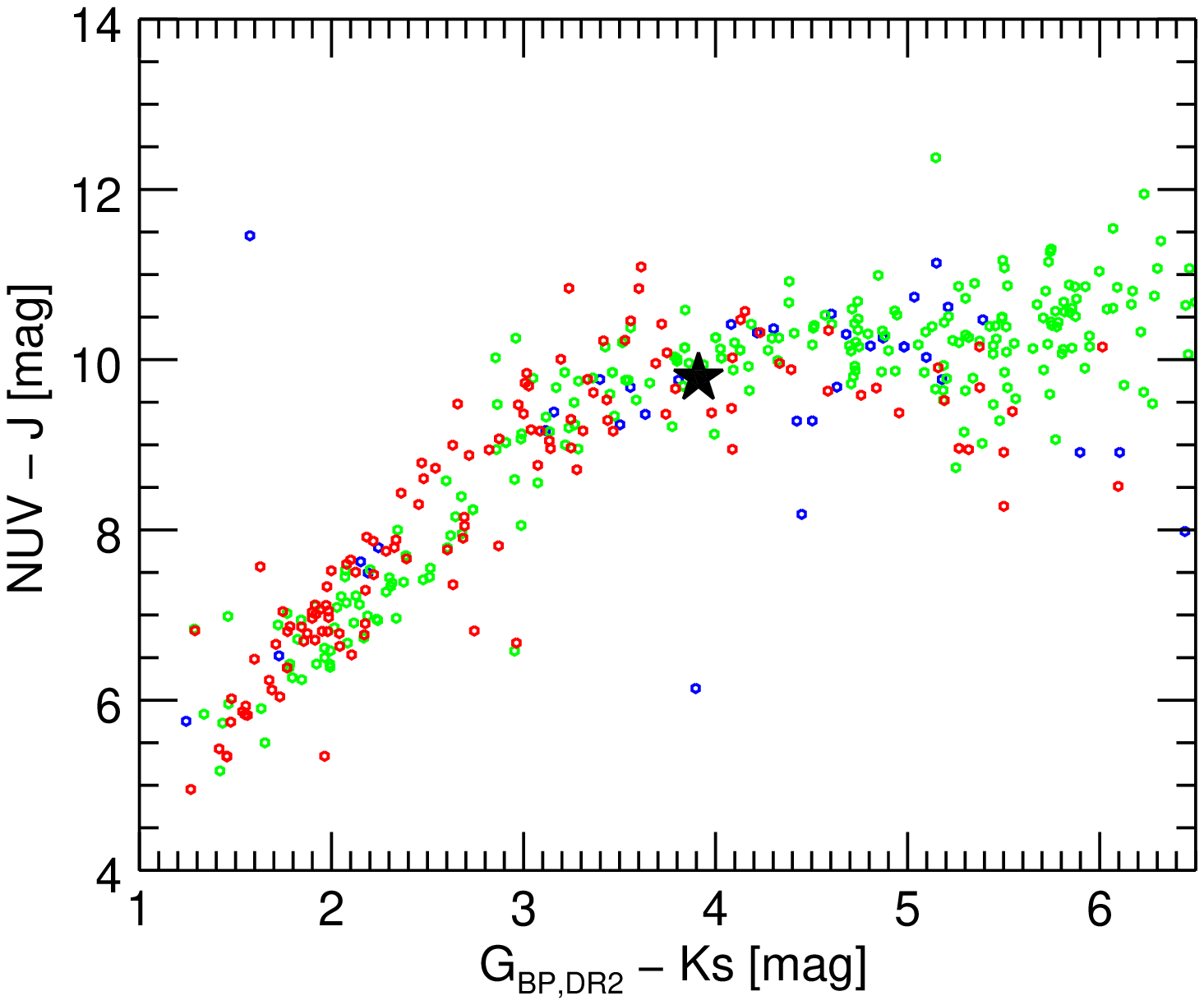}
     \end{minipage}
     \begin{minipage}[htbp]{0.33\hsize}
       \includegraphics[width=\hsize]{\dirGJ1108A/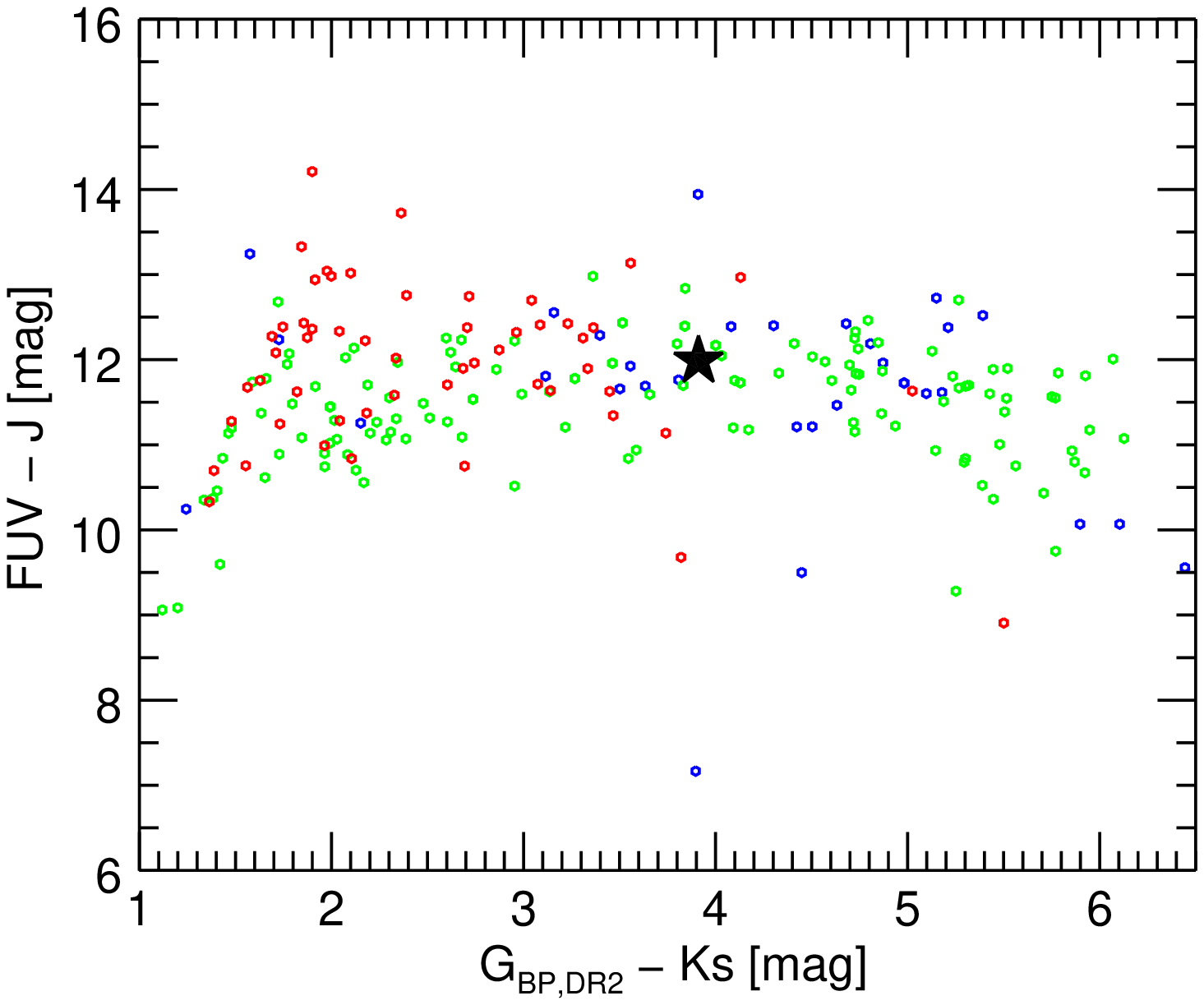}
     \end{minipage} \\

     \begin{minipage}[htbp]{0.33\hsize}
       \includegraphics[width=\hsize]{\dirGJ1108A/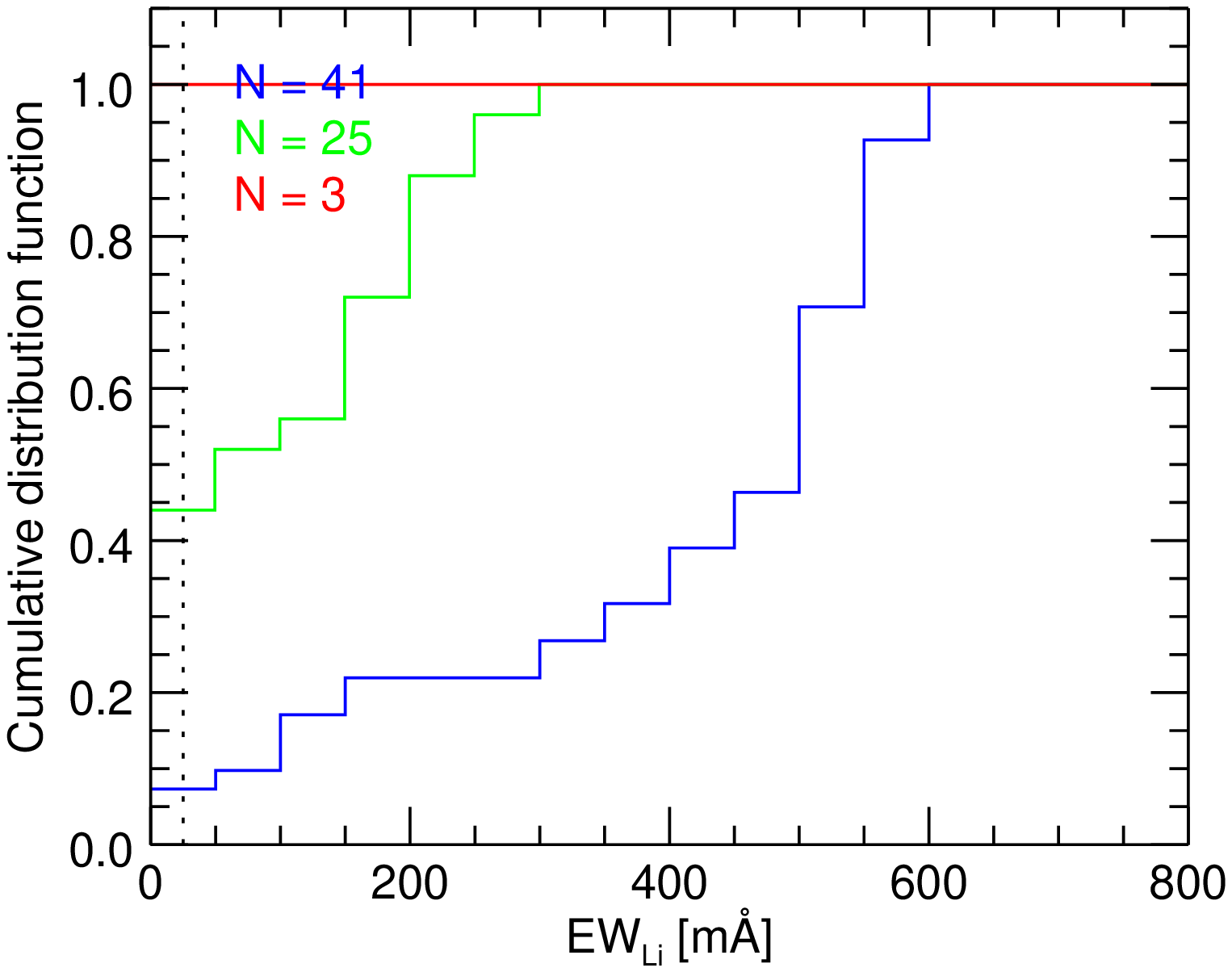}
     \end{minipage}
     \begin{minipage}[htbp]{0.33\hsize}
       \includegraphics[width=\hsize]{\dirGJ1108A/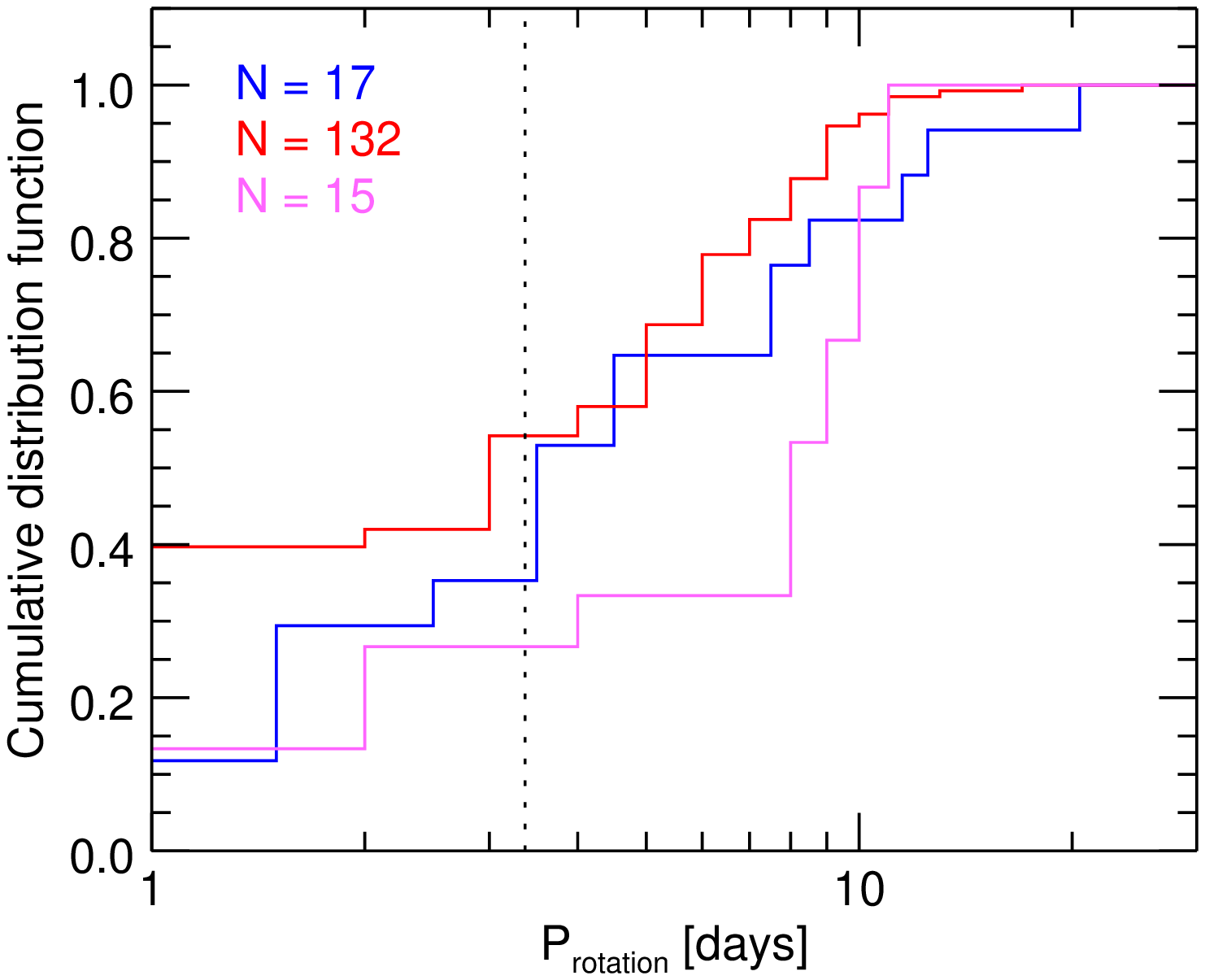}
     \end{minipage}
   \end{tabular}
  \caption{Age indicators based on activities are represented. From the top left to middle right; equivalent width of lithium, spin period, equivalent width of H$\alpha$, UV excesses are presented. The filled star indicate GJ1108A, and stars in several young groups are also shown with according colors; blue for TW Hydrae, $\beta$ Pictoris, $\gamma$ Velorum; green for Carina, Columba, Tucana-Horologium, and Octans; red for AB Dorudos, Pleiades, M35; and pink for M34 (See the text for references). It should be noted that stars in those panels may not be common; some of them have investigated by only imaging or spectroscopic observations. In the bottom, cumulative distribution function of lithium equivalent width and spin period are shown, in which GJ1108A is represented as a dashed line.}
  \label{fig:age_indicators}
\end{figure*}

\subsection{Age constraint using indicators} \label{subsec:indicator}

Young low-mass stars display several indicators of their youth and magnetic activity \citep[e.g.][]{skumanich72}. In the early stage of stellar evolution, strong lithium absorption at 6708 $\AA$ is observed in the atmospheres of low-mass stars \citep[e.g.][]{mentuch08, dasilva09}. As PMS stars evolve and contract, their cores become hotter and reach a temperature where the hydrogen burning can occur, a few million degrees Kelvin depending on density and metallicity. In the P-P chain nuclear reaction, lithium is consumed/deplete and therefore only young stars and objects not massive enough to stably fuse hydrogen, such as brown dwarfs, have lithium in their atmospheres. Lithium depletion is an important age indicator that may be less dependent on the effects of rotation and magnetic fields, and has been used to discover new members of nearby moving groups \citep[e.g.][]{murphylawson15}, though the depletion timescale has spreads due to other effects, such as episodic accretion \citep{baraffe10}. GJ1108A has little lithium in its atmosphere \citep{lopezsantiago10}. \cite{brandt14a} reports an equivalent width upper limit of 25 m$\AA$, indicating that it is not very young.

In general, the stellar spin rate is considered to correlate with age. If the presence of a circumstellar disk is not considered, pre-main-sequence stars spin up as they contract, until hydrogen burning is ignited. Once the zero-age-main-sequence phase is reached, angular momentum is lost to the stellar wind through so-called magnetic braking \citep[e.g.][]{weberdavis67}. Several effects induced by the magnetic field can also be seen in the form of chromospheric activity through line emission at optical wavelengths, or coronal activity through X-ray emission, which can be used as empirical age indicators. Calibrated relations for the age as function of rotational rate and $B-V$ color (so-called gyrochronology) have been presented in \cite{barnes07} and \cite{mamajekhillenbrand08}. The rotational period of GJ1108A is determined using light curves from the HATNet transit survey for exoplanets, yielding $\tau_{\rm rot}=3.37$ days \citep{hartman11}, and the color $B-V=1.28$ is provided by the Tycho-2 catalogue \citep{hog00}. However, the gyrochronology method is poorly calibrated for such red and very active stars as GJ1108A, and thus cannot produce a conclusive age estimate \citep{mamajekhillenbrand08}. Similarly, it is difficult to determine the age for such stars with chromospheric activity indicators.

X-ray activity indicators are relatively well applicable to characterize the age of red and active stars. The X-ray excess is coming from coronal heating coupled with magnetic activity, which can be seen in young stars and represented as the ratio to the bolometric luminosity, log$(L_{\rm X}/L_{\rm bol})=R_{\rm X}$. The correlation between $R_{\rm X}$ and age (or chromospheric activities) for solar-type stars has been established \cite[e.g.][]{mamajekhillenbrand08}. The relation for late-type stars is less clear, because the X-ray luminosity of late-type stars does not decay as steeply with age as in solar-type stars \citep{preibischfeigelson05}, and the X-ray index of stars younger than $\sim100$ Myr saturates around log$R_{\rm X}=-3.0$ \citep{pizzolato03}. \cite{jackson12} determined the isochrones in a wider range of $B-V$ (mag) and $R_{\rm X}$, providing the ages where X-ray luminosity saturates as log$(\tau_{\rm sat})=8.21\pm0.31$ yr with $R_{\rm X}=-3.14\pm0.02(0.35)$ dex for $B-V$=1.275--1.410 stars. The $R_{\rm X}$ of GJ1108A system is indeed large: $-3.13$ dex, as obtained from the ROSAT All Sky Survey \citep{voges99}. Thus we may be able to set $\tau_{\rm sat}=162.181_{-82.75}^{+168.95}$ Myr as an upper limit for the age of GJ1108A, which is consistent with the other indications. However, it should be noted that the X-ray luminosity of GJ1108Aa might be overestimated due to the existence of GJ1108Ab and GJ1108B.

In order to understand chromospheric activity and stellar age, we use UV fluxes taken by $GALEX$ \citep{bianchi14} because of its good sensitivity for nearby stars and higher spatial resolution than $ROSAT$. The UV activities also saturate at young age of about 100 Myr \citep{shkolnikbarman14,ansdell15}, the trend of which is similar to the coronal activity observed in X-ray. In UV-NIR color-color diagrams, young stars tend to show blue colors \citep[e.g.][]{bowler12b}. In $GALEX$ images, GJ1108A is less-contaminated by GJ1108B due to their large separation, and the UV-NIR colors of the target is consistent with those in nearby stellar group younger than 200 Myr (The middle two panels of Figure \ref{fig:age_indicators}).

\begin{deluxetable*}{ccc ccc ccc c}
\centering
\tablewidth{0pt}
\tablecaption{Observing Log of imaging}
\tablehead{
    \colhead{Obs. Date (UT)} &
    \colhead{Instrument} &
    \colhead{$N_{\rm{exp}}$ } &
    \colhead{$t_{\rm{tot}}$} &
    \colhead{Filter} &
    \colhead{Airmass} &
    \colhead{$m_{\rm prim}$} & 
    \colhead{$m_{\rm comp}$} &
    \colhead{Position angle} &
    \colhead{Separation} \\
    \colhead{yyyy-mm-dd} & \colhead{} & \colhead{} & \colhead{[sec]} & \colhead{} &
    \colhead{} & \colhead{[mag]} & \colhead{[mag]} & \colhead{[degree]} & \colhead{[arcsec]}
}
\startdata
2001-11-30 & NIRC2 & 4 & 20 & PK50$\_$1.5+$NB2.108$ & 1.03 & - & - & 50.96$\pm$0.21 & 0.154$\pm$0.002 \\
2004-04-04 & NIRC2 & 8 & 32 & PK50$\_$1.5+$NB2.108$ & 1.05 & - & - & 86.17$\pm$0.27 & 0.264$\pm$0.002 \\
2011-12-25 & HiCIAO & 30 & 45  & $H+$ND10    & 1.04 & 6.74$\pm$0.04 & 8.73$\pm$0.04 & 80.89$\pm$0.53 & 0.253$\pm$0.002 \\
2013-01-01 & HiCIAO & 20 & 30  & $J+$ND10    & 1.05 & 7.37$\pm$0.02 & 9.31$\pm$0.02 & - & - \\
           &        & 10 & 15  & $H+$ND10    & 1.05 & 6.74$\pm$0.02 & 8.73$\pm$0.04 & 91.17$\pm$0.13 & 0.267$\pm$0.001 \\
           &        & 10 & 15  & $K_{s}$+ND10 & 1.05 & 6.55$\pm$0.03 & 8.55$\pm$0.03 & - & - \\
2014-01-21 & HiCIAO & 50 & 75  & $H$+ND1     & 1.79 & 6.74$\pm$0.02 & 8.72$\pm$0.02 & 102.16$\pm$1.03 & 0.249$\pm$0.003 \\
2015-01-07 & HiCIAO & 90 & 450 & $H$+ND1     & 1.75 & 6.73$\pm$0.02 & 8.79$\pm$0.02 & 115.02$\pm$0.31 & 0.202$\pm$0.001 \\
2015-12-29 & HiCIAO & 10 & 40  & $J$+ND1     & 1.13 & 7.36$\pm$0.02 & 9.38$\pm$0.02 & - & - \\
           &        & 30 & 150 & $H$+ND1     & 1.11 & 6.74$\pm$0.02 & 8.74$\pm$0.02 & 142.20$\pm$0.34 & 0.114$\pm$0.001 
\enddata
\label{tab:imglog} 
\end{deluxetable*} 

\begin{deluxetable*}{ccc ccc cc}
\centering
\tablewidth{0pt}
\tablecaption{Observing Log of echelle spectroscopy}
\tablehead{
    \colhead{Obs. Date (UT)} &
    \colhead{Instrument} &
    \colhead{$\lambda$} &
    \colhead{$\lambda$/$\Delta\lambda$} &
    \colhead{$t_{\rm{exp}}$} &
    \colhead{Airmass} &
    \colhead{BERV} &
    \colhead{RV} \\
    \colhead{yyyy-mm-dd} & \colhead{} & \colhead{[nm]} & \colhead{} & \colhead{[sec]} &
    \colhead{} & \colhead{[km]} & \colhead{[km]}
    }
\startdata
2007-11-25 & SOPHIE & 387--694  & 75000  & 1800 & 1.02 & 23.71 &  9.48$\pm$0.23 \\
2007-11-30 & SOPHIE & 387--694  & 75000  & 1800 & 1.02 & 22.18 &  9.60$\pm$0.23 \\
2012-11-20 & HDS    & 497--758  & 110000 & 600  & 1.04 & 25.19 & 12.14$\pm$0.45 \\
2012-11-26 & ARCES  & 320--1000 & 31500  & 1000 & 1.03 & 23.55 & 10.74$\pm$0.60 \\
2015-11-28 & HDS    & 489--769  & 70000  & 240  & 1.05 & 23.11 & 11.01$\pm$0.56 \\
2016-10-14 & HDS    & 489--769  & 70000  & 240  & 1.09 & 29.41 &  4.83$\pm$0.52
\enddata
\label{tab:rvlog}
\end{deluxetable*}

To evaluate activities of GJ1108A and constrain its age, GJ1108A is compared with objects in several young stellar groups. We compiled young stars including candidates and their observed properties from the literature \citep[][and references therein]{ansdell15, dasilva09, frasca15, kraus14, meibom09, meibom11, messina17a, newton17, rebull16, schneider12}. The compiled young stars are categorized into four groups; young population (5--30 Myr) including TW Hydrae, $\beta$ Pictoris, $\gamma$ Velorum; intermediate (30--60 Myr) with Carina, Columba, Tucana-Horologium, and Octans; old (100--200 Myr) with AB Dorudos, Pleiades, M35; and M34 (220 Myr). Their colors are determined by cross-identification with 2MASS PSC \citep{skrutskie06} and $Gaia$ Data Release 2 \citep{gaiadr2}. We here employ the $G_{\rm BP}$ magnitude of $Gaia$ to understand optical flux of M-dwarfs since $Gaia$'s high sensitivity with very wide FoV coverage is important to characterize many further and faint M-dwarfs.

We here combine all the age indicators of GJ1108A to understand its age. The lithium depletion can place a constraint on the lower limit for stellar age. As seen in top left of Figure \ref{fig:age_indicators} \citep[See also][]{murphylawson15}, all the object in young population have significant lithium absorption, down to $\sim$200 m$\AA$. GJ1108A has just an upper limit for lithium equivalent width of 25 m$\AA$ \citep{brandt14a}, and therefore the system is probably older than stars in TW Hydrae association, $\sim$20 Myr (in the top left of Figure \ref{fig:age_indicators}, almost all the stars in young population belong to TW Hydrae association). Gyrochronology has large spread of the spin-age relation for young red stars, especially at redder than $G_{\rm BP,DR2},-K_{s}\sim4$ [mag]. We described the cumulative distribution function for stars with $G_{\rm BP,DR2}-K_{s}=(G_{\rm BP,DR2}-K_{s})_{\rm GJ1108A}\pm0.3$, and found that many stars ($\sim55\%$) in the old population (100--200 Myr) have shorter spin periods than GJ1108A but just $30\%$ of stars in M34 (220 Myr) do, indicating 220 Myr may be a marginal upper limit for the age of GJ1108A. Chromospheric activity seen in UV excess and accretion signature of H$_{\alpha}$ emission (6563 $\AA$) are not adequate to accurately estimate stellar age for low-mass stars, although they are useful to distinguish young stars from old stars. We do not employ X-ray activity as an age indicator for GJ1108A, since the photometry by $ROSAT$ is contaminated not only by GJ1108Ab but also widely separated system GJ1108B due to its poor spatial resolution. In summary with current knowledge of indicators for GJ1108A, independently from kinematics, the target should be in the age range of 20--220 Myr.


\section{Observation and Analysis}
\indent

\subsection{Subaru/HiCIAO}
\begin{figure*}
  \centering
  \begin{tabular}{cccc}
     \begin{minipage}[htbp]{0.23\hsize}
       \includegraphics[width=\hsize]{\dirGJ1108A/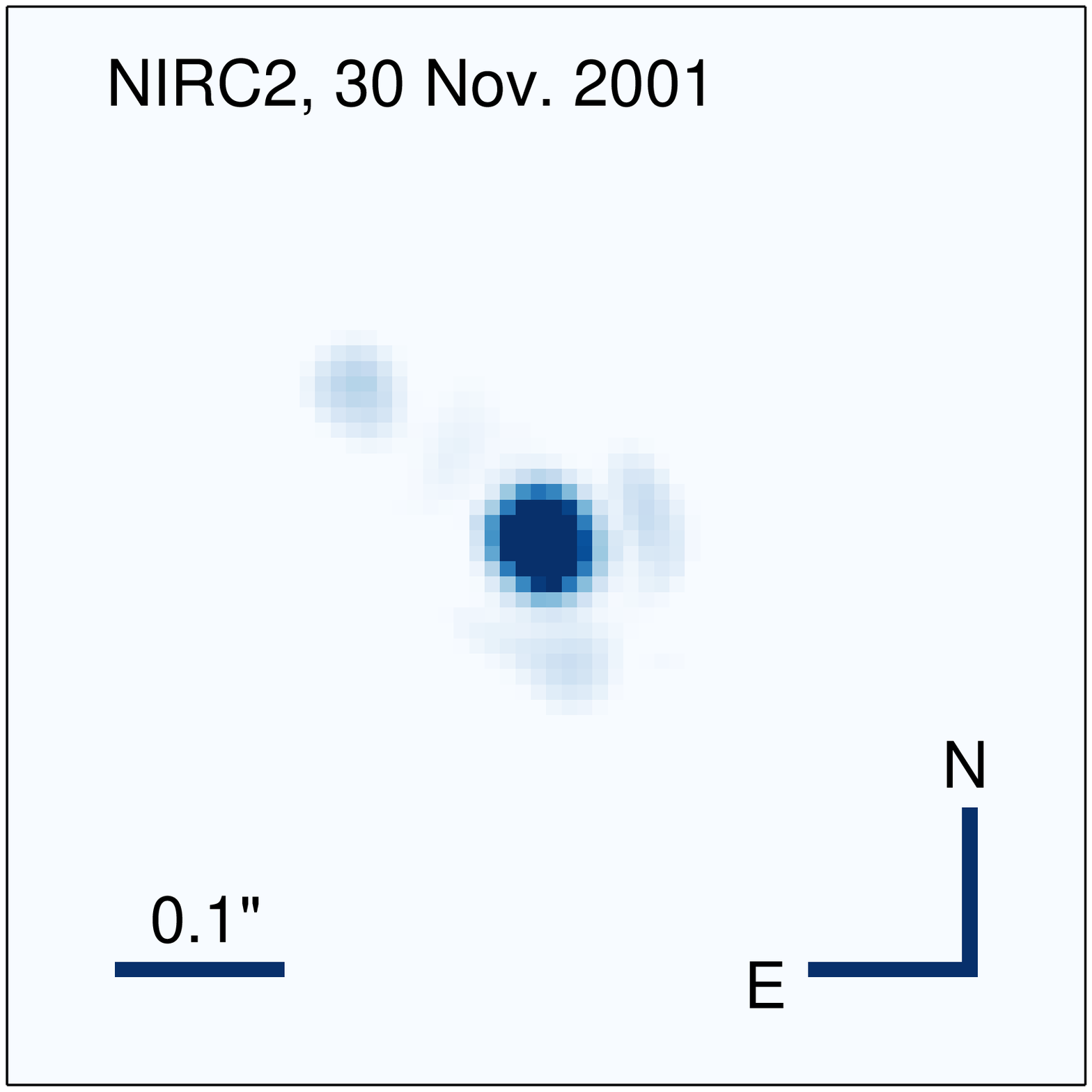}
     \end{minipage}
     \begin{minipage}[htbp]{0.23\hsize}
       \includegraphics[width=\hsize]{\dirGJ1108A/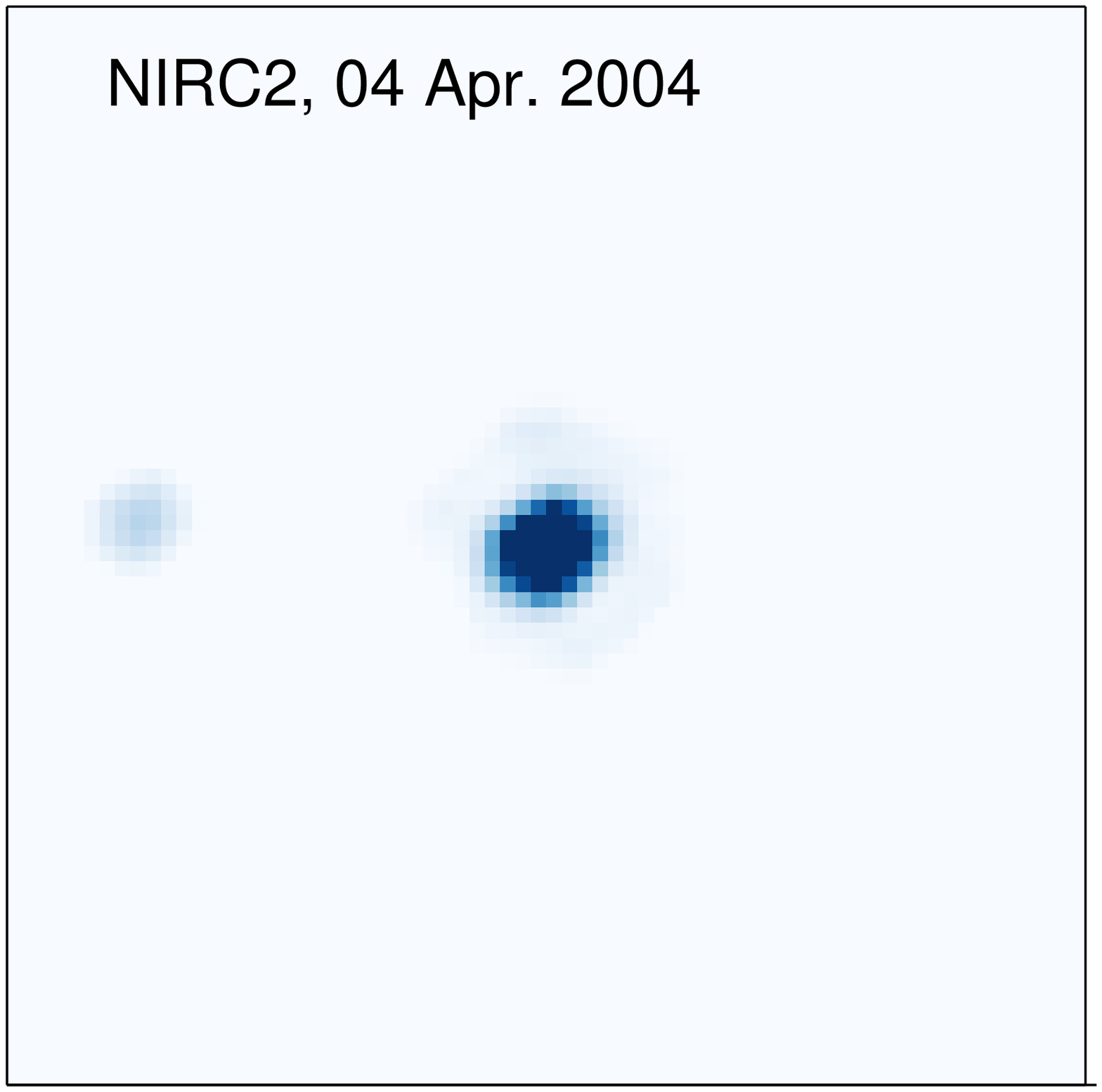}
     \end{minipage}
     \begin{minipage}[htbp]{0.23\hsize}
       \includegraphics[width=\hsize]{\dirGJ1108A/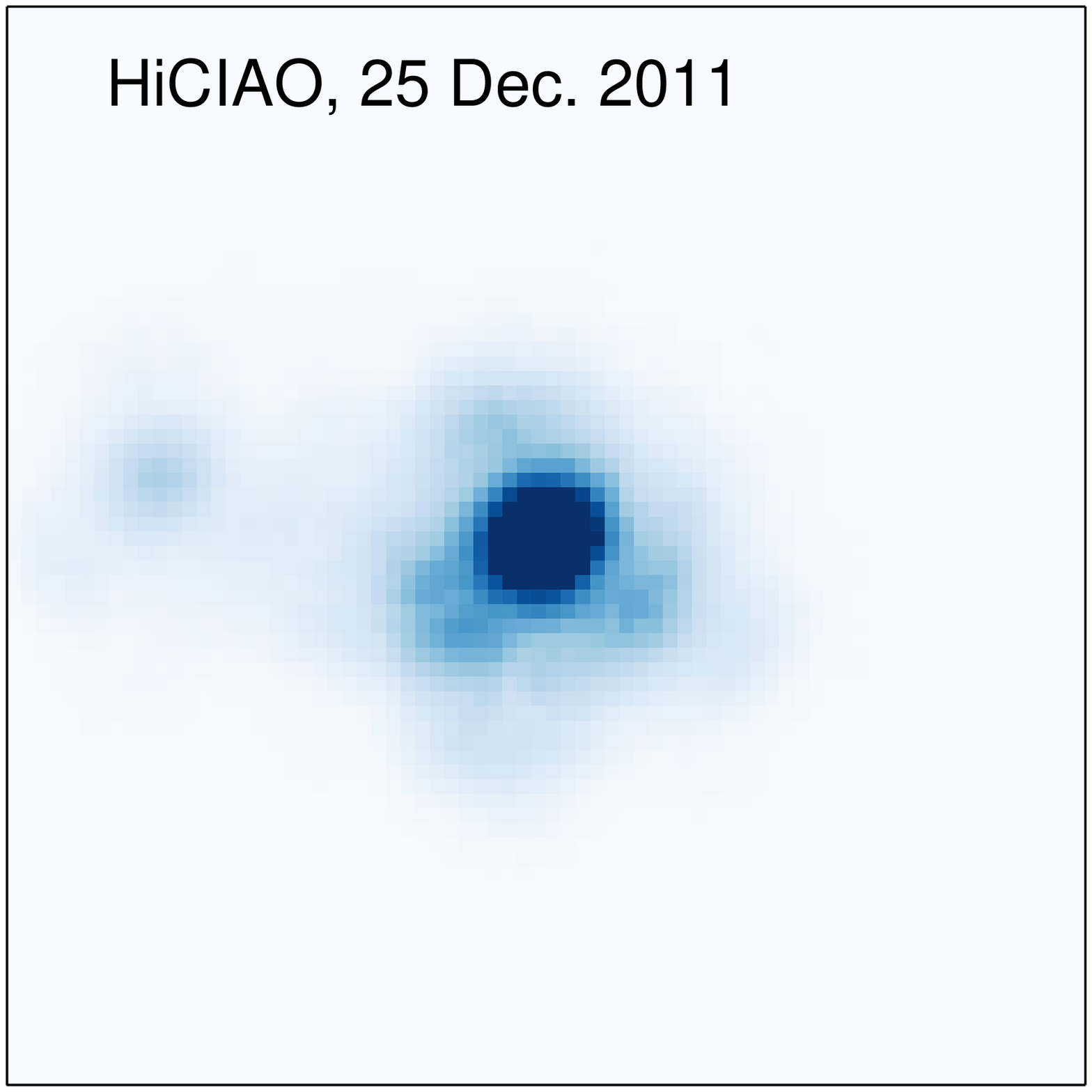}
     \end{minipage}
     \begin{minipage}[htbp]{0.23\hsize}
       \includegraphics[width=\hsize]{\dirGJ1108A/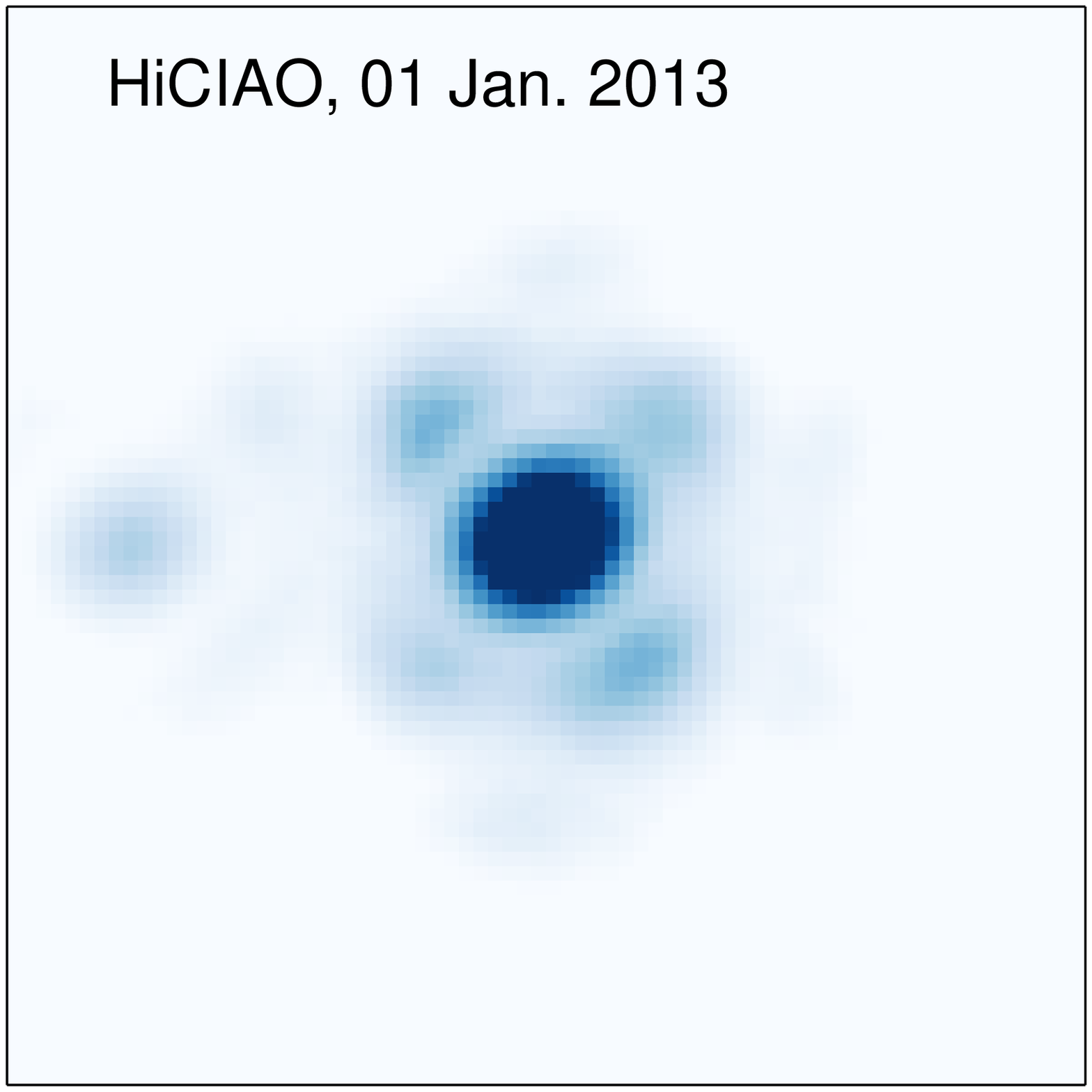}
     \end{minipage} \\
     \begin{minipage}[htbp]{0.23\hsize}
       \includegraphics[width=\hsize]{\dirGJ1108A/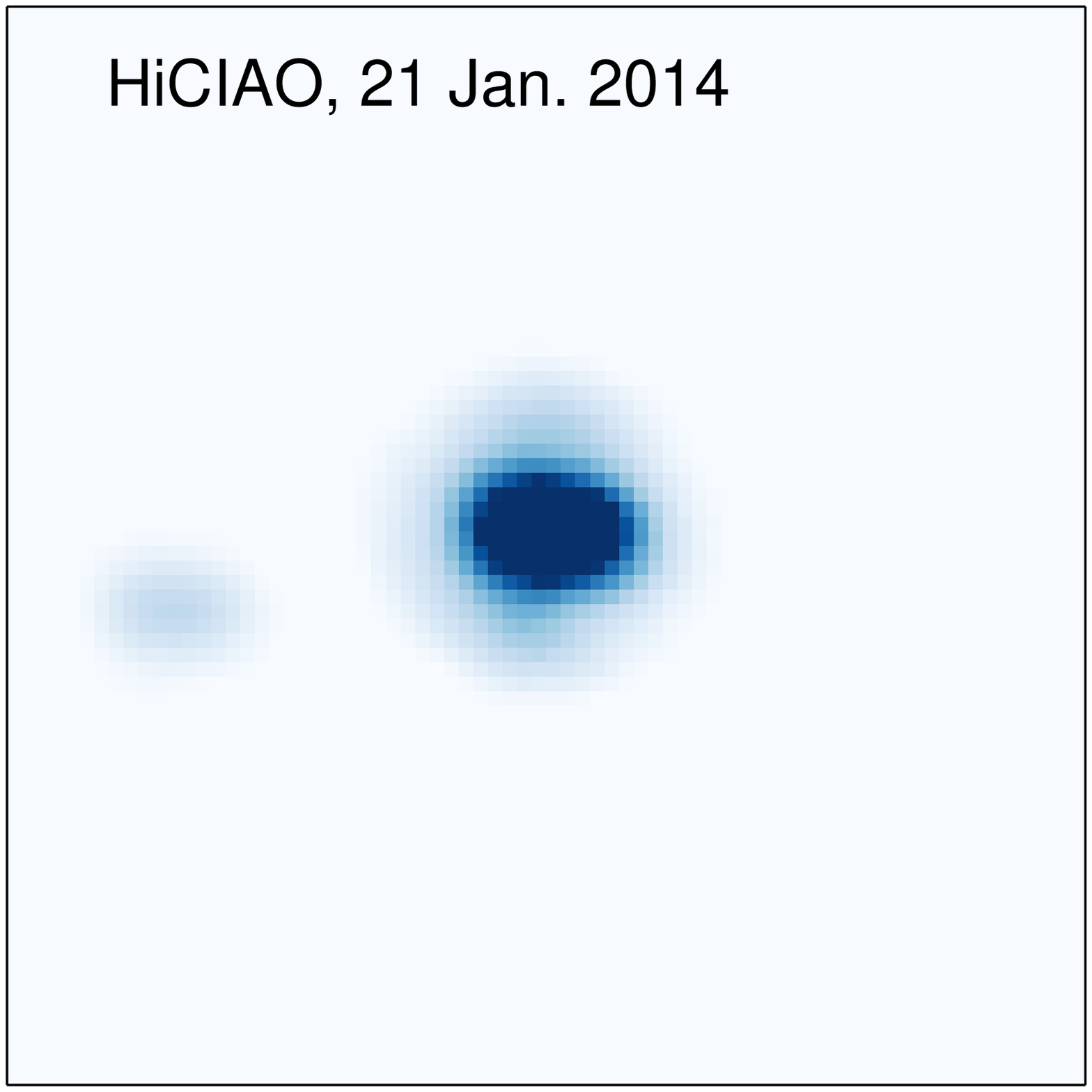}
     \end{minipage}
     \begin{minipage}[htbp]{0.23\hsize}
       \includegraphics[width=\hsize]{\dirGJ1108A/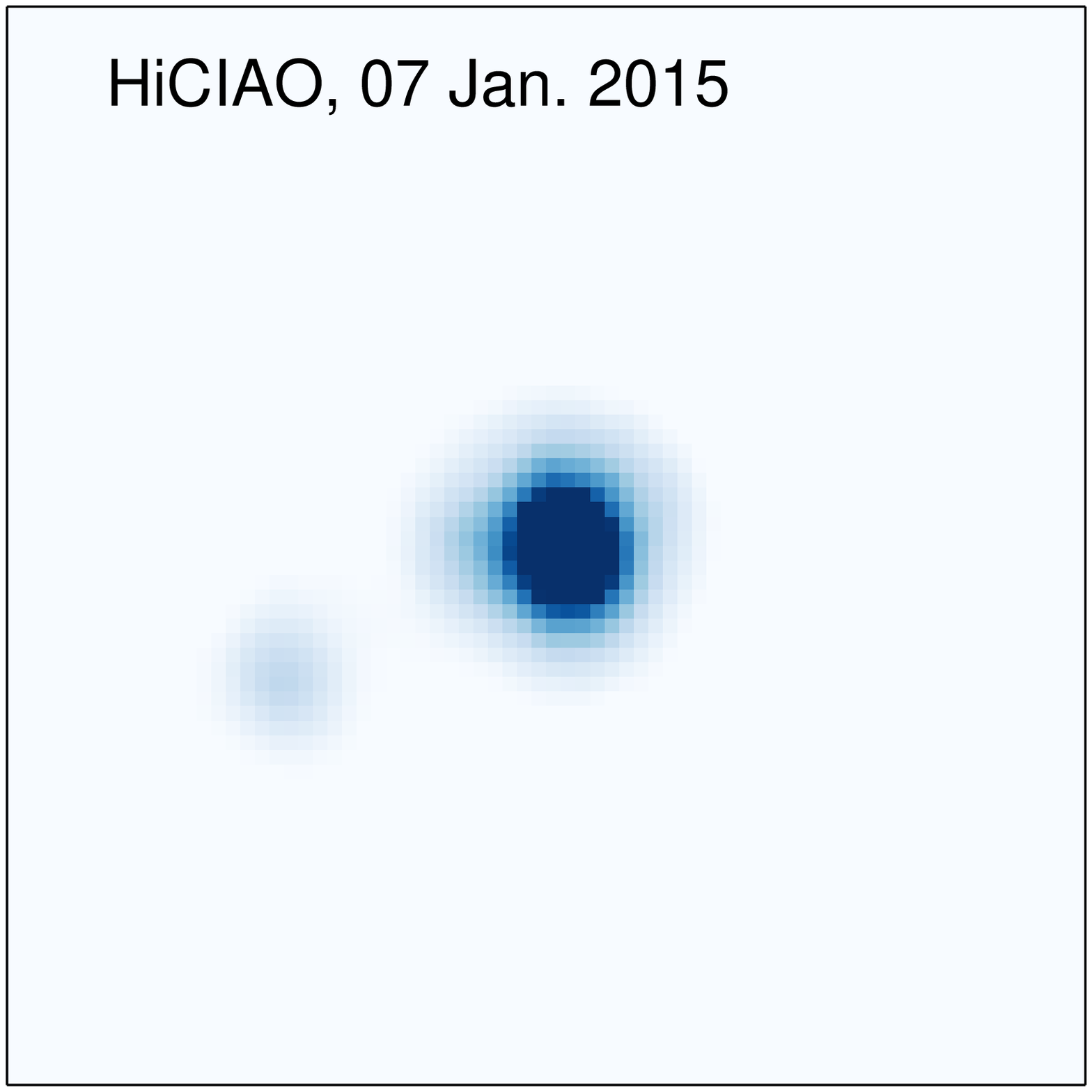}
     \end{minipage} 
     \begin{minipage}[htbp]{0.23\hsize}
       \includegraphics[width=\hsize]{\dirGJ1108A/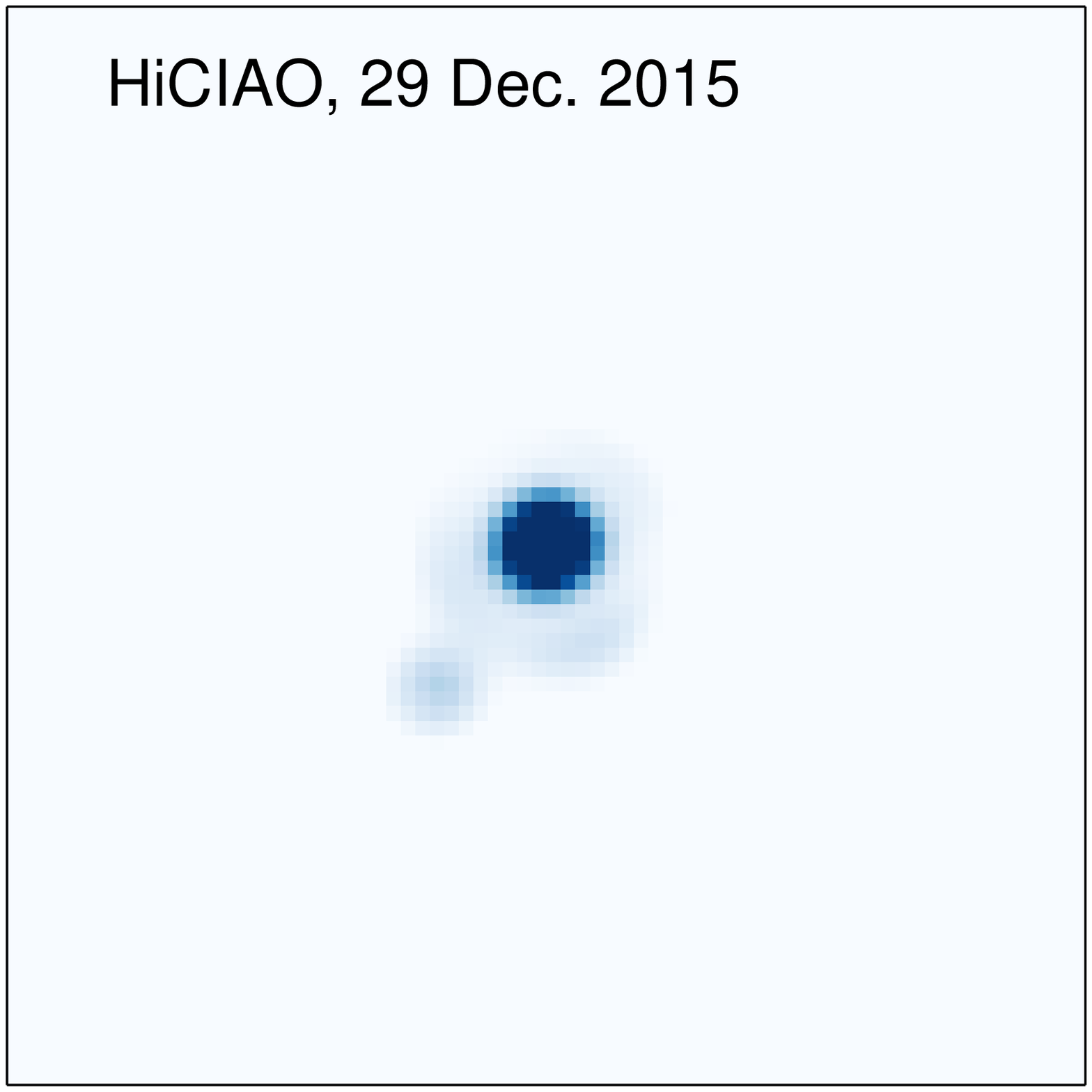}
     \end{minipage}
   \end{tabular}
  \caption{High-resolution images of GJ1108A system taken by NIRC2 and HiCIAO. All images are aligned such that North is up and East is to the left, with a $0.7\arcsec$ field of view. Structures like point source may be seen around the primary. We conclude those are local speckle or diffraction patterns because similar patterns are also found around the companion and they have not been found by the follow-up observations.}
  \label{fig:images}
\end{figure*}

We observed GJ1108A as part of the SEEDS project, which aims to improve the understanding of the formation and evolution of massive planets and disks with ages mainly in the range of 1.0~Myr to 1.0~Gyr. Observations were conducted with a combination of the HiCIAO camera \citep{Suzuki10} and AO188 \citep{hayano08, hayano10} adaptive optics system. Neutral density filters were used to enable precise photometry and astrometry for each object. The observing parameters of the imaging are summarized in Table \ref{tab:imglog}, and reduced images are presented in Figure \ref{fig:images}.

As a first step in our reduction procedure for imaging data from HiCIAO, the detector stripe pattern is modelled in sky regions and then subtracted. Flat fielding and deviant pixel correction are then performed for the de-striped images. Distortion in the HiCIAO camera, plate scales of each axis on the detector, and an offset of position angle are calibrated using an MCMC approach for the M5 and M15 globular cluster \citep{brandt13, helminiak16}. The precision of the HiCIAO's distortion corrections is provided in \citet{helminiak16}.

A photometry of the companion requires removal of photometric contamination from the primary, since GJ1108Ab is separated by just $\sim0.17\arcsec$ from the primary and their contrast is about a factor of 6.4 \citep{brandt14a}. We attempted rejecting the contamination as follows. First, each frame is convolved with the photometric aperture, r=0.5FWHM. The sigma-clipped mean of the counts at the separation of the companion in the convolved image is used to estimate the level of contamination from the primary. This is then subtracted from the photometric count at the specific location of the companion. In order to calibrate the total system brightness, the 2MASS photometry of the unresolved GJ1108A system was used \citep{skrutskie06}. 

For astrometry, its error budget is not dominated by photon noise in cases where both a primary and its companion are bright. The AO-corrected PSF of high-resolution imaging has fast-changing asymmetric structures, so-called speckle patterns, which are difficult to properly model. The structure tends to dominate the noise budget over the Poisson noise of photon counting. Meanwhile, since the companion is very close to the primary and their contrast is also very small, the separation with standard PSF fitting tends to be underestimated. To estimate the separation of the binary, instead of registering the center of brightness distribution, we searched the centroid in a narrow area with a radius of FWHM from their peak location, corresponding to PSF core for each object. The PSF shape of both the primary and the companion should be almost same (Figure \ref{fig:images}), and hence the separation between each PSF core should be the separation of the system. We measured the separations between the primary and secondary stars on each reduced image, providing an averaged separation as a final measurement of astrometry and a standard deviation that represents the error bar of each astrometry.

\subsection{KeckII/NIRC2}
The GJ1108A system was observed by KeckII/NIRC2 with a narrow-band filter in $K$-band in 2001 and 2004, and we obtained the corresponding raw images from the Keck Observatory Archive\footnote{https://koa.ipac.caltech.edu/cgi-bin/KOA/nph-KOAlogin}. The initial image processing consisted of dark subtraction, flat fielding, deviant pixel correction, and sky subtraction. The distortion solution in \cite{yelda10} was applied to images, enabling the suppression of the instrumental errors in astrometry. Additionally, the offset of position angle given on the NIRC2 website\footnote{https://www2.keck.hawaii.edu/inst/nirc2/nirc2$\_$ao.html$\#$pa} was also corrected. We did not conducted photometry for those data taken by NIRC2 with the narrow-band filter, because we do not know a photometric magnitude of GJ1108A system at the narrow-band filter.

\begin{figure*}[htbp]
  \centering
  \begin{tabular}{cc}
     \begin{minipage}[t]{0.45\hsize}
       \includegraphics[width=\hsize]{\dirGJ1108A/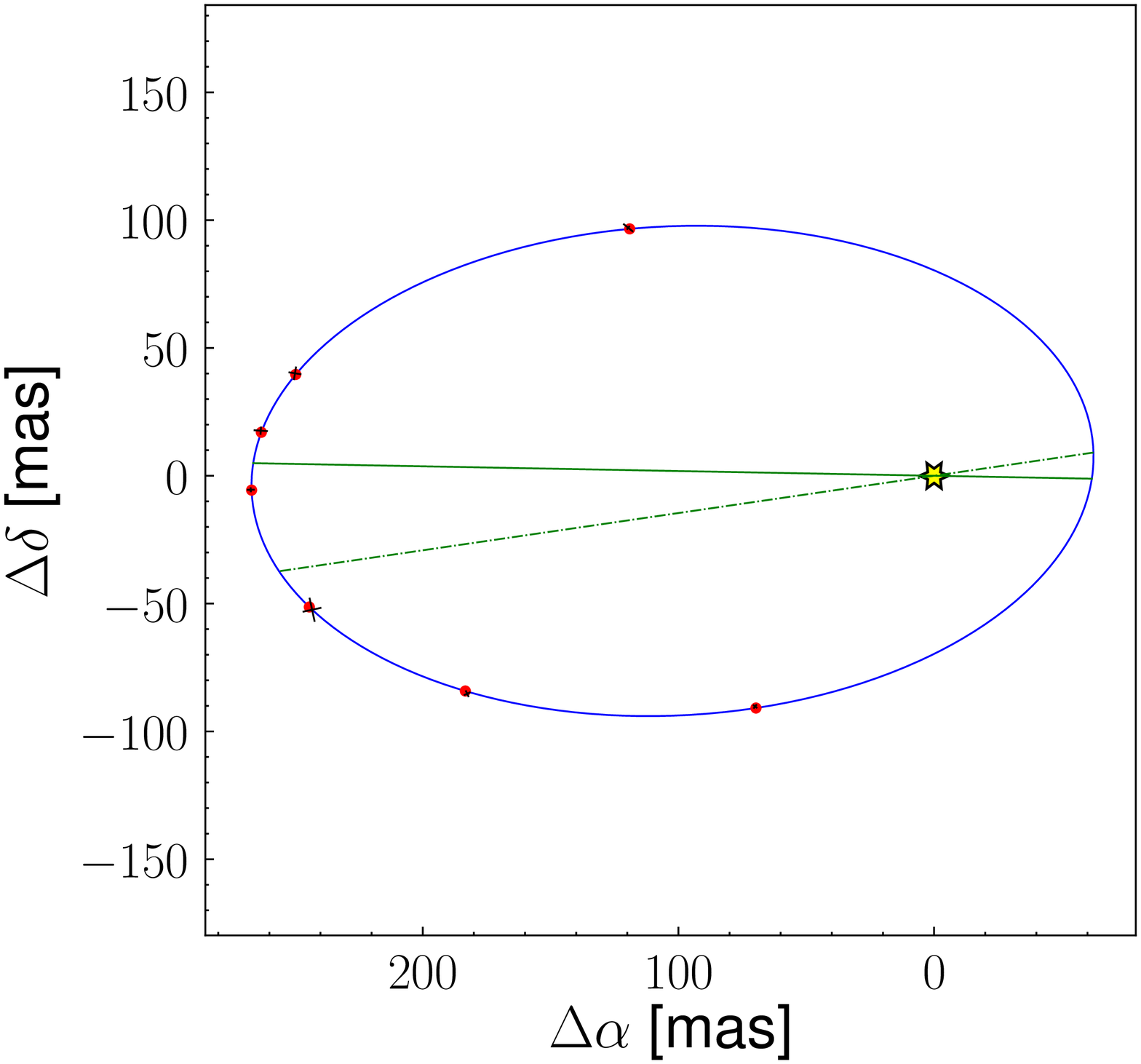}
     \end{minipage}
     \begin{minipage}[t]{0.45\hsize}
       \includegraphics[width=\hsize]{\dirGJ1108A/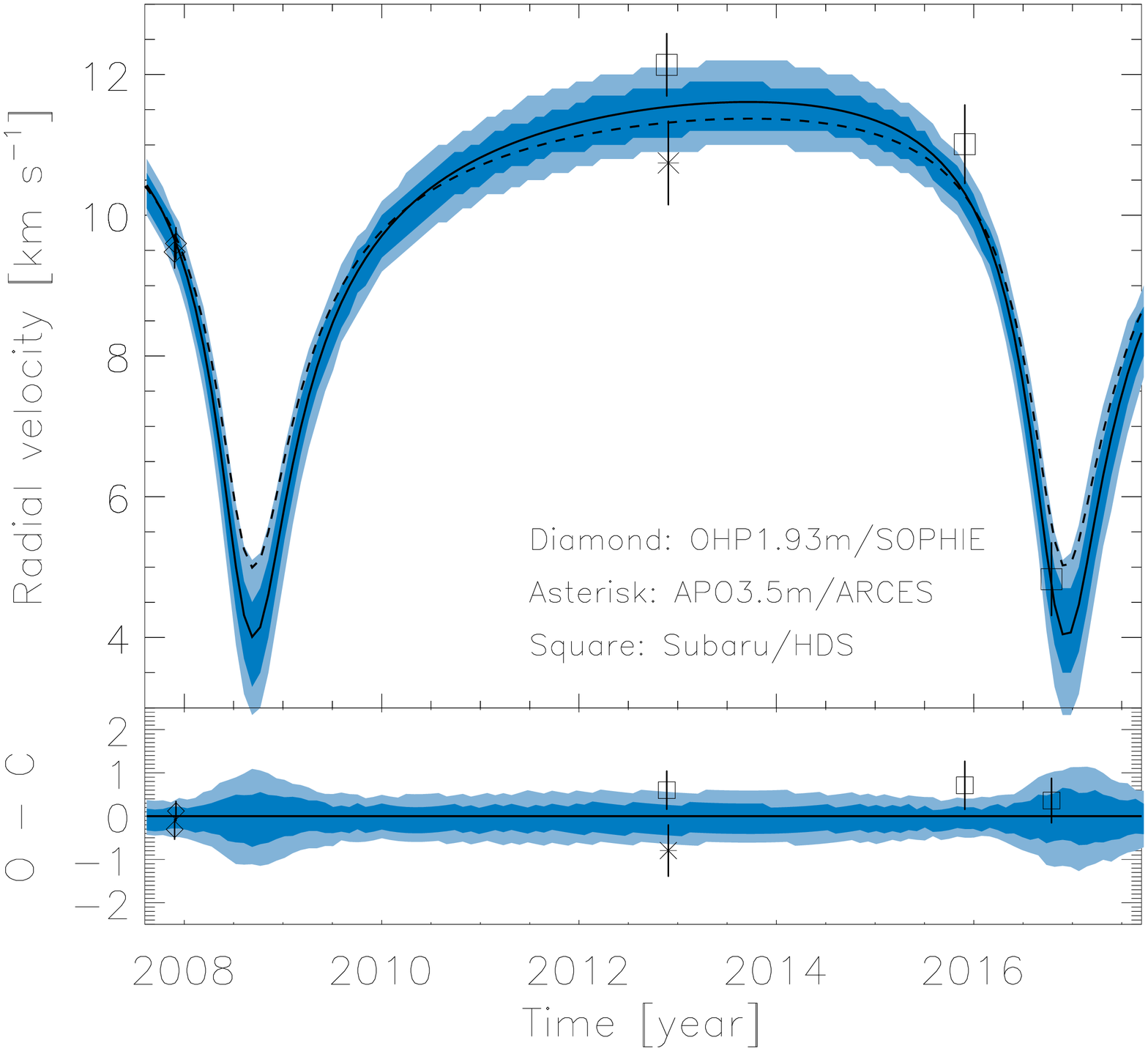}
     \end{minipage}
   \end{tabular}
  \caption{The best-fit relative orbit of GJ1108Ab around GJ1108Aa is shown to the left. The star symbol, solid line, and dashed-dotted line indicate the position of primary, line of apsides, and line of nodes, respectively. The panel to the right shows the most probable RV curve of GJ1108A, in which blue regions indicate confidence range of 68 and 95$\%$ and the dashed line represents the RV curve calculated with some of orbital elements from our astrometry analysis and the mass estimates that are based on the models of \cite{baraffe15}.
  }
\end{figure*}

\subsection{Radial velocity measurements}
The SOPHIE archive\footnote{http://atlas.obs-hp.fr/sophie/} has two datasets of GJ1108A, which had already passed their image reduction procedure, wavelength calibration using simultaneous Th-Ar lamp, and Barycentric Earth Radial Velocity correction. We downloaded those archival data.

We observed GJ1108A with the Astrophysical Research Consortium Echelle Spectrograph (ARCES) on the Astrophysical Research Consortium 3.5 meter telescope at the Apache Point Observatory (APO) \citep{wang03} on 2012 November 26. The strength of the system's Li 6708 $\AA$ absorption line derived from these data was previously reported in \cite{brandt14a}. ARCES provided a R $\sim$ 31500 spectra that covered the wavelength range of 3500 $\AA$ to 10200 $\AA$. Long-term monitoring of a known radial velocity standard has revealed that ARCES has an RV stability of $\sim$0.5 km s$^{-1}$ \citep{mack13}. Our 1000 second integration yielded a signal to noise ratio (SNR) of $\sim$90 near 6050 $\AA$. We reduced these data using standard techniques in IRAF \footnote{IRAF is distributed by the National Optical Astronomy Observatory, which is operated by the Association of Universities for Research in Astronomy (AURA) under a cooperative agreement with the National Science Foundation.}. We utilized a Thorium-Argon lamp exposure taken immediately after our observation of GJ1108A to perform wavelength calibration on these data, and also applied standard heliocentric velocity corrections.

We also observed the object with HDS \citep{noguchi02} on 20 November 2012, 28 November 2015, and 14 October 2016. One-dimensional spectra were obtained with IRAF procedures following a standard manner of HDS data reduction including bias subtraction, flat fielding, subtraction of scattered light, and wavelength calibration using emission lines of the Thorium-Argon lamp non-simultaneously taken on the same nights. Observing properties of high-dispersion spectroscopy are summarized in Table \ref{tab:rvlog}.

In order to understand the orbital RV modulation by combining all the datasets for GJ1108A, we recalculated helio- and barycentric radial velocity for each data point with a single manner and measured the shift of telluric absorption lines to compensate the zero-point offsets of each facility. The re-wavelength calibration was performed with a typical telluric template \citep{bertaux14}. We used iSpec \citep{blancocuaresma14} for general analyses including the wavelength calibration and radial velocity measurements. Since telluric spectra may drift \citep{gray06, figueira10c}, we verified the RV stability by applying our RV-measurement method to datasets of an RV standard star, HD9407, taken with SOPHIE. The same zero-point correction was applied to the standard star, and we found the RVs of the standard star were indeed unstable by 0.1 km s$^{-1}$ as an RV scatter from a mean center of a zero-point offset. On the other hand, in the case of HDS with $\tau$ Cet, the wavelength re-calibration using telluric absorption has a larger uncertainty, corresponding to 0.4 km s$^{-1}$. The difference to SOPHIE may be because SOPHIE has advantages in RV determination; the spectrograph is temperature-controlled and simultaneous wavelength calibration is performed. Those scatter are adopted as a typical uncertainty of zero-point correction using telluric spectra.

For the reduced spectra taken by ARCES and HDS, their RV measurements were performed as follows. Orders of an one-dimensional spectrum were grouped into three segments. For each segment, we measured a shift to a K5 template mask equipped in iSpec via cross-correlation analysis. The mask size is determined by the rotational velocity of GJ1108A, $v$sin$i$ $\sim$ 12.5 km s$^{-1}$ taken from the SOPHIE archive. We adopted the velocity steps in calculating the cross-correlation function was a tenth of a template mask's size. The same approach was also employed for the zero-point correction using telluric absorption with a telluric template and calculating steps depending on instrumental spectral resolutions. The cross-correlation function of GJ1108A had single peak. This is because the contrast in optical wavelength should be larger than in infrared, about a few tens time (See Appendix \ref{sec:app}), and a slit configuration might enhance the contrast in the slit. The final RV measurements and their errors for the ARCES and HDS data were obtained by calculating an average and a standard deviation of the RV calculations in three segments. In the case of SOPHIE archival data, we employed the RV uncertainty calculated by iSpec. It should be noted that those uncertainty, potentially coming from photon noise, do not dominate error budget of RV. Since the GJ1108Aa is a rapidly rotating star, we add 168 m s$^{-1}$ of an uncertainty potentially induced by the stellar activity \citep{bailey12} in the final error values, as well as the errors of RV measurements and their offset calibrations.


\begin{deluxetable}{ccc}
\centering
\tablewidth{0pt}
\tablecaption{Bayesian priors used in the ExoSOFT}
\tablehead{ \colhead{Parameter} & \colhead{Prior} & \colhead{Range}}
\startdata
            $m_{1} \, \& \, m_{2}$ ($M_{\odot}$) & PDMF\tablenotemark{a}   & 0.08-1.0 \\
            $\varpi$ (mas) & Gaussian\tablenotemark{b} $\times$ (1/$\varpi^{4}$) & 30--65 \\
            $e$              & $p(e) = 2e$                        & 0--0.98 \\
            $P$ (yr)         & Power-law\tablenotemark{c}         & 0--30 \\
            $i$ (degree)     & $p(i) \propto$ sin($i$)            & 0--180  \\
            $\Omega \, \& \, \omega$ (degree) & Uniform & 0--360 \\
            $\gamma$ (km/s)  & Uniform                            & 0--20 
\enddata
\tablenotetext{a}{Present-Day Mass Function \citep{chabrier03}}
\tablenotetext{b}{The parallax is referred from \cite{vanleeuwen07}.}
\tablenotetext{c}{\cite{cumming08}}
\label{tab:priors}
\end{deluxetable}

\section{Orbital solution}
\indent

The general form of Kepler's third law may be written as
\begin{eqnarray}
  a_{1}+a_{2} = \left[ \frac{P^{2}G \left(m_{1}+m_{2}\right)}{4\pi^{2}} \right]^{1/3}
  \label{eq:kepler3_temp}
\end{eqnarray}
where the $a_{1} \ a_{2}, \, P, \, G, \ m_{1}$, and $m_{2}$ respectively indicate semimajor axis of primary, semimajor axis of companion, orbital period, gravitational constant, mass of primary, and mass of companion. In cases where the barycenter of a binary system can be measured with wide field-of-view observations using telescopes such as $Hipparcos$ and $Gaia$, $a_{1}$ and $a_{2}$ are separately obtained. However the barycenter cannot be obtained with high-resolution AO-imaging observations due to a very narrow field-of-view. Alternatively, those observations precisely estimate the separation of binary systems. For the case of GJ1108A system, separation were typically obtained with several tenth's of an arcsecond accuracy (Table \ref{tab:imglog}). For high-resolution AO-imagings, the Eq. \ref{eq:kepler3_temp} is rewritten as
\begin{eqnarray}
  a_{\rm total} = \left[ \frac{P^{2}G m_{\rm total}}{4\pi^{2}} \right]^{1/3}
  \label{eq:kepler3}
\end{eqnarray}
where $a_{\rm total}$ and $m_{\rm total}$ indicate separation of a binary system and total mass of the system, respectively. Although true orbits are not determined by only high-resolution imaging, the equation can be solved without knowing a barycenter of a system: motions due to their proper motion in the Galaxy and annual parallax. In order to differentiate $m_{\rm total}$ into its $m_{1}$ and $m_{2}$ components, radial velocity measurements are required.

\begin{deluxetable}{cc}[t]
\centering
\tablewidth{0pt}
\tablecaption{Results of orbital fit}
\tablehead{ \colhead{Parameter} & \colhead{PDMF(Uniform)} }
\startdata
$m_{1} (M_{\odot})$          &   $   0.72_{  -0.04}^{+  0.04}$   ($   0.72_{  -0.04}^{+  0.04}$) \\
$m_{2} (M_{\odot})$          &   $   0.30_{  -0.03}^{+  0.03}$   ($   0.30_{  -0.03}^{+  0.03}$) \\
$\varpi$ (mas)             &     $  40.27_{  -0.36}^{+  0.36}$   ($  40.27_{  -0.36}^{+  0.36}$) \\
$\Omega (^{\circ})$         &    $ 278.22_{  -1.77}^{+  1.72}$   ($ 278.22_{  -1.75}^{+  1.75}$) \\
$e$                        &     $   0.63_{  -0.01}^{+  0.01}$   ($   0.63_{  -0.01}^{+  0.01}$) \\
$T_{0}$ (JD+2454703)        &    $   0.85_{  -9.88}^{+ 10.04}$   ($   0.90_{  -9.99}^{+  9.99}$) \\
$P$ (yr)                   &     $   8.24_{  -0.03}^{+  0.03}$   ($   8.24_{  -0.03}^{+  0.03}$) \\
$i (^{\circ})$              &    $  42.47_{  -1.21}^{+  1.32}$   ($  42.46_{  -1.22}^{+  1.31}$) \\
$\omega (^{\circ})$         &    $ 347.45_{  -2.58}^{+  2.60}$   ($ 347.45_{  -2.62}^{+  2.57}$) \\
$a_{\rm total}$ (AU)         &   $   4.11_{  -0.05}^{+  0.05}$   ($   4.11_{  -0.05}^{+  0.05}$) \\
$K$ (km/s)                 &     $   3.80_{  -0.37}^{+  0.37}$   ($   3.80_{  -0.37}^{+  0.37}$) \\
$\gamma$ (km/s)            &     $  10.14_{  -0.17}^{+  0.17}$   ($  10.14_{  -0.17}^{+  0.17}$) \\
$\chi^{2}_{\rm 3D}$(best-fit)  &  0.89 (0.88) \\
$\chi^{2}_{\rm astr}$(best-fit) &  0.39 (0.41) \\
$\chi^{2}_{\rm RV}$(best-fit)  &  5.76 (5.54) \\
\hline
$a_{\rm total}$ ($\arcsec$)  & 0.166$\pm$0.001 \\
$m_{\rm total} (M_{\odot})$ & 1.033$\left(\frac{d}{24.832 {\rm pc}}\right)^{3}\pm m_{\rm total}\sqrt{0.0007+9{\left( \frac{\Delta d}{d} \right)}^{2}}$ \\
$m_{1} (M_{\odot})$ & $m_{\rm total}$ - $m_{2}$ \\
$m_{2} (M_{\odot})$ & 0.297$\left(\frac{K}{3.800 {\rm km/s}}\right) \times {m_{\rm total}}^{2/3}$ 
\enddata
\label{tab:orbparam}
\end{deluxetable}

\begin{figure*}
  \centering
  \includegraphics[width=\hsize]{\dirGJ1108A/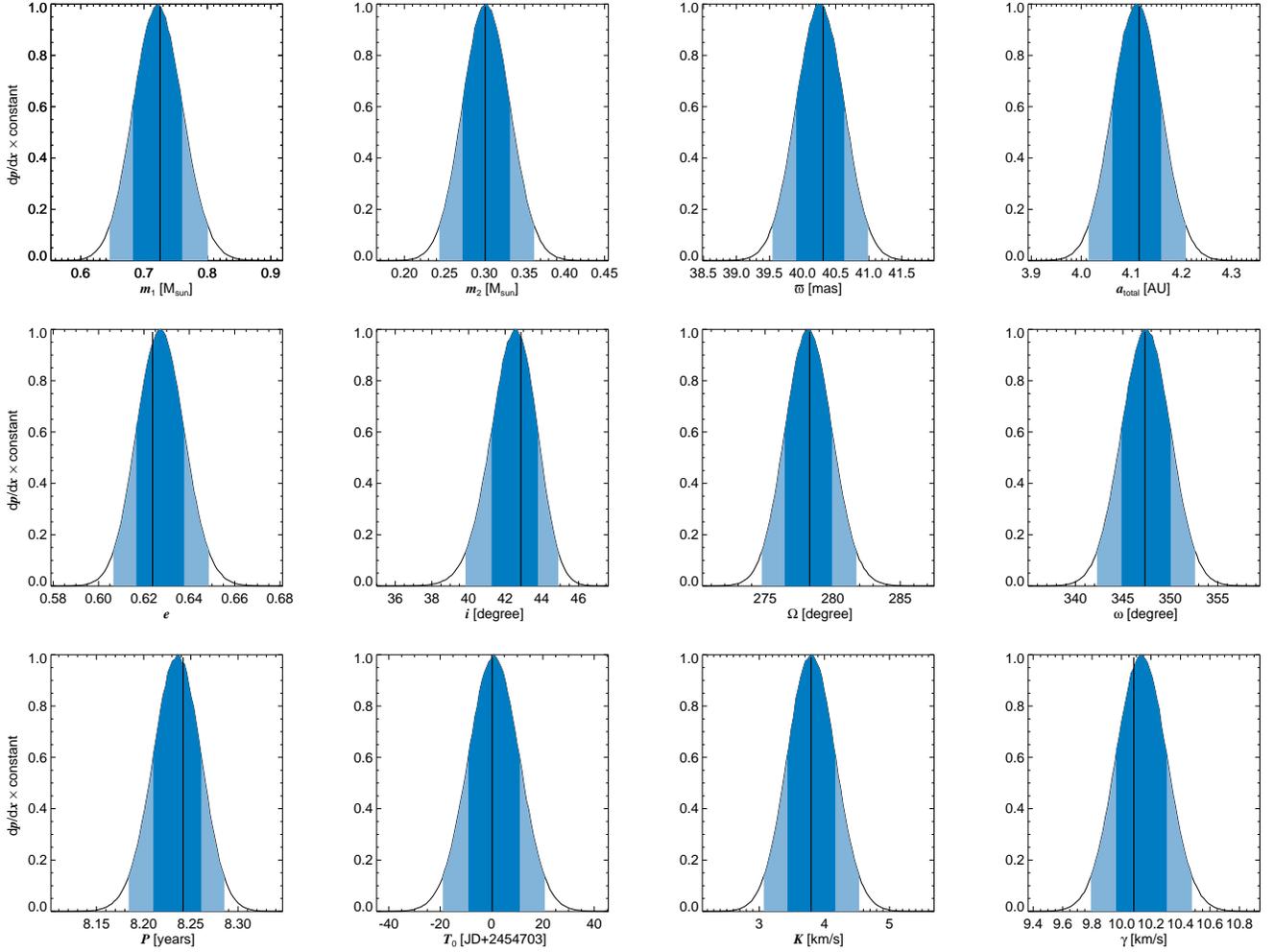}
  \caption{Normalized probability distributions of each orbital element. Black vertical lines indicate the best-fit parameters. The blue regions show confidence ranges of 68 and 95$\%$.}
\end{figure*}

\begin{figure*}[t]
  \centering
  \begin{tabular}{ccc}
     \begin{minipage}[htbp]{0.30\hsize}
       \includegraphics[width=\hsize]{\dirGJ1108A/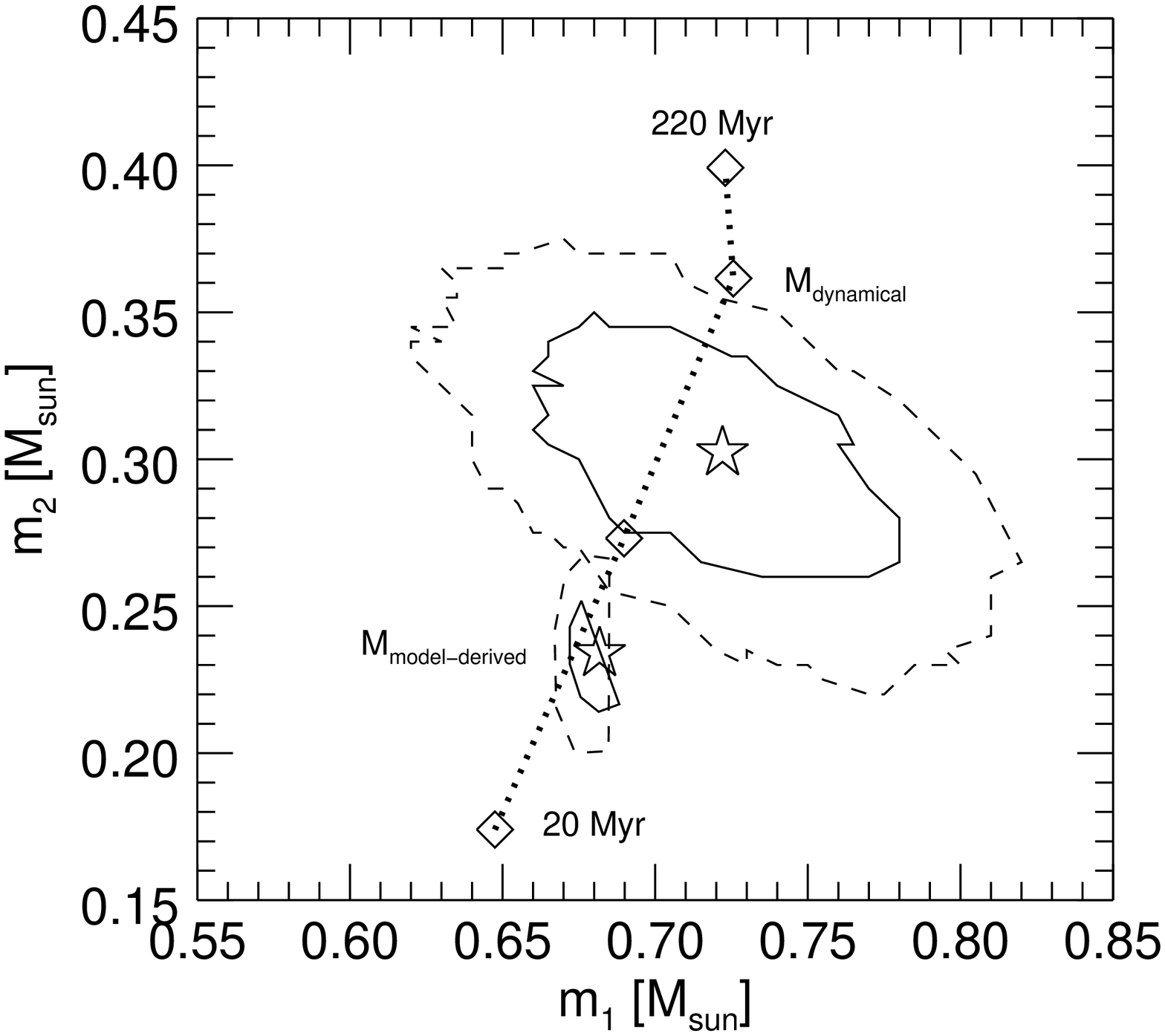}
      \end{minipage}
     \begin{minipage}[htbp]{0.30\hsize}
       \includegraphics[width=\hsize]{\dirGJ1108A/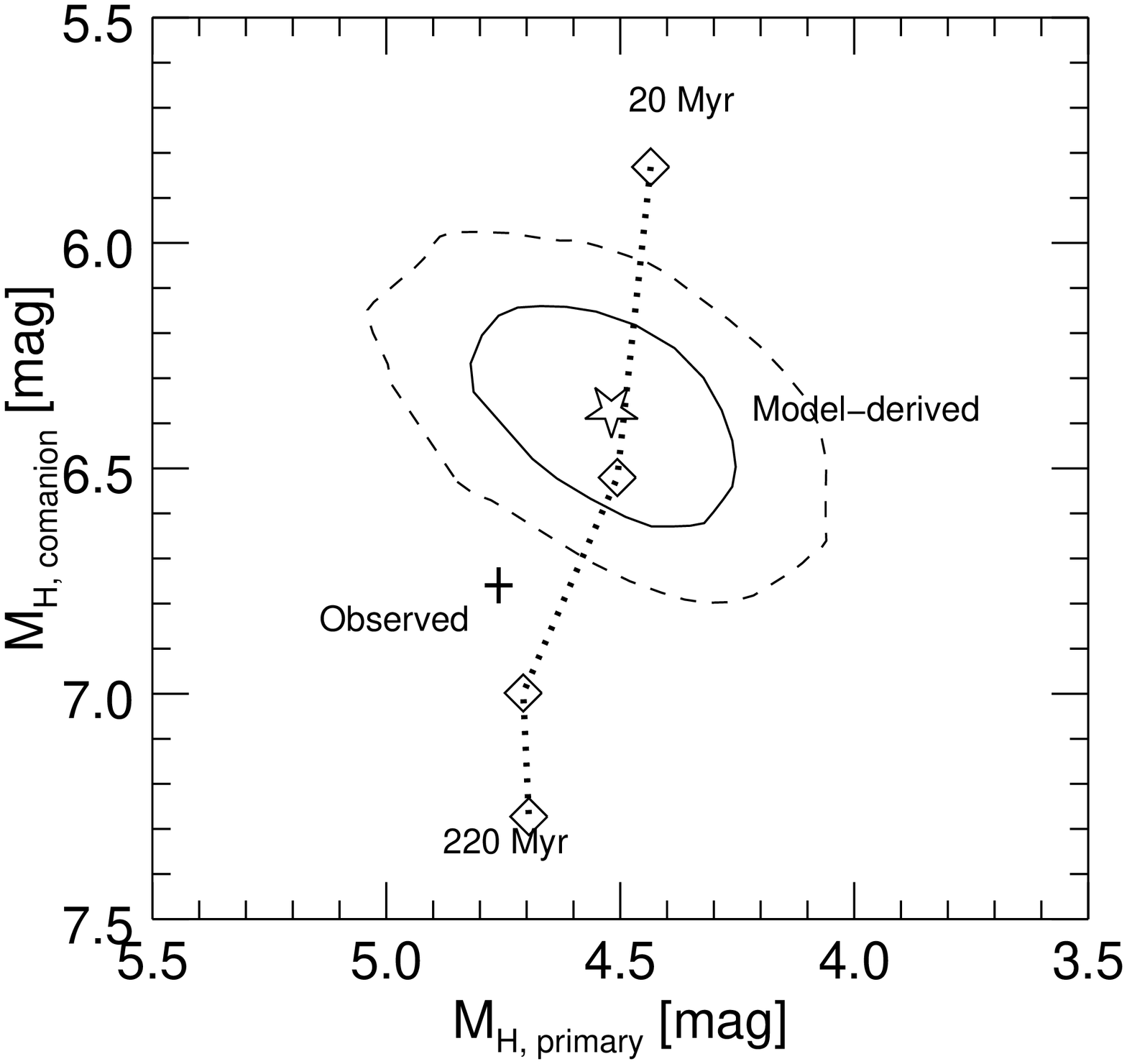}
     \end{minipage}
     \begin{minipage}[htbp]{0.30\hsize}
       \includegraphics[width=\hsize]{\dirGJ1108A/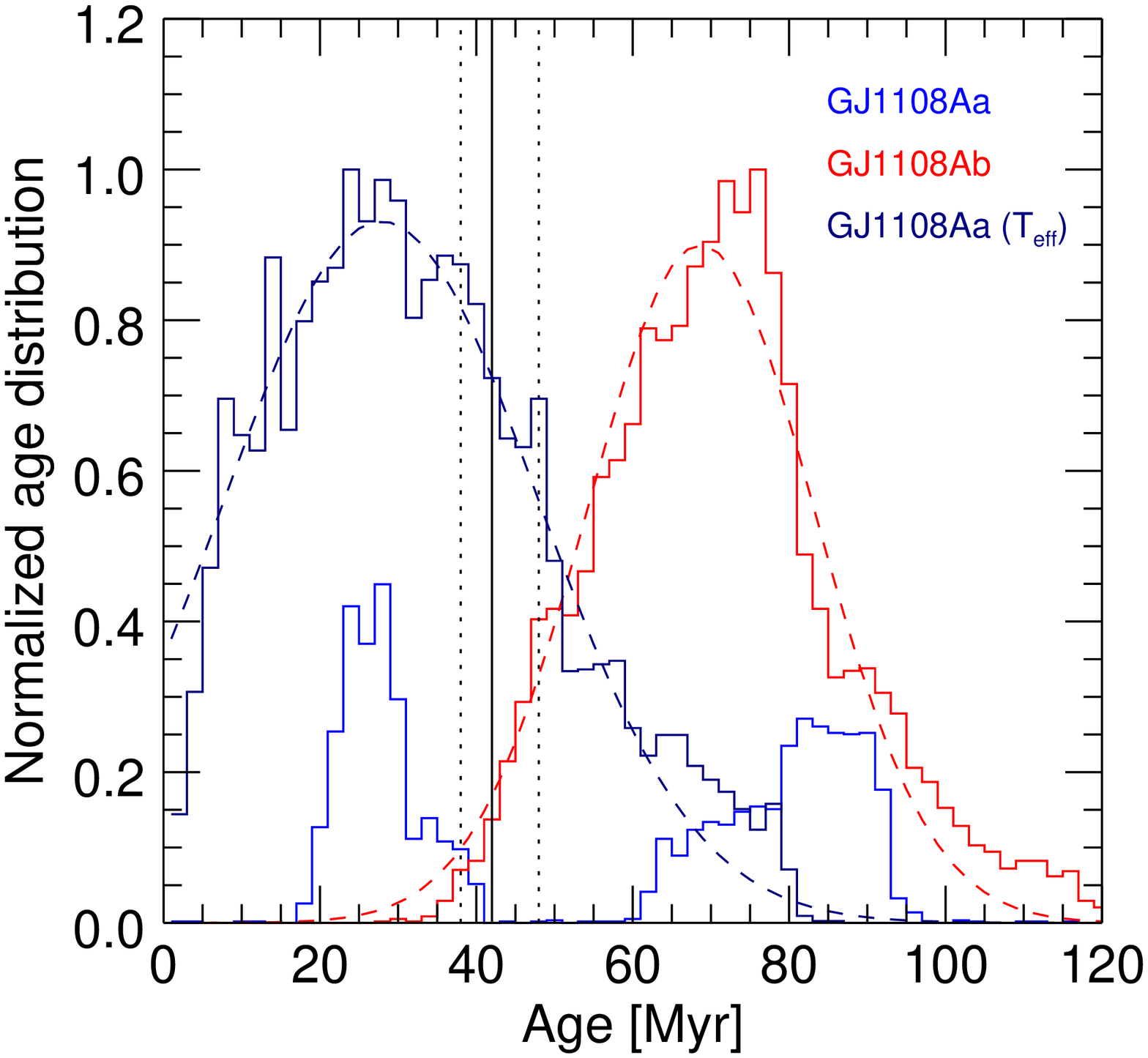}
     \end{minipage} 
   \end{tabular}
  \caption{The left panel shows contour maps for the mass of GJ1108Aab derived from two different approaches. The solid and dashed present 68 and 95$\%$ contours, respectively. In the middle panel, observed $H$-band flux and the model-derived flux are compared. In these panels, the age of Columba is assumed for GJ1108A system. Diamonds with dotted lines assumes the age range suggested by indicators; 20, 50, 100, and 220 Myr. The right panel shows model-derived age distributions of the primary and the companion with the age of Columba presented by black lines. To derived parameters based on BHAC15 evolutionary model in these panels, we randomly sampled the input variables with Monte-Carlo simulation.}
  \label{fig:masscomp_gj1108A}
\end{figure*}

For three-dimensional orbital fitting, we used the Exoplanet Simple Orbit Fitting Toolbox; ExoSOFT \citep[][https://pypi.python.org/pypi/ExoSOFT]{mede17}. The code employs various Bayesian fitting approaches, including the Markov Chain Monte Carlo, to obtain an orbital solution with observed properties of astrometry and RV. The Keplerian model for the three dimensional orbital fitting in the tool has main parameters: mass of each component, orbital eccentricity, orbital period, parallax, time of last periapsis, longitude of ascending node, argument of periapsis, separation of binary, and radial velocity of a system due to its proper motion along a line of sight which are respectively presented as: $m_{1}, \ m_{2}, \ e, \ P, \ \varpi, \ T_{0}, \ i, \ \Omega, \ \omega, \ a_{\rm total}$, and $\gamma$. 

From several fitting options in ExoSOFT, we employed SAemcee mode to find an orbital solution, in which solutions are initially searched with simulated annealing, and then MCMC via emcee \citep{foremanmackey13} on the basis of the optimization found by the simulated annealing stage. Each parameter's search range is provided in Table \ref{tab:priors}. We ran 4 MCMC simulations in parallel, each of which has $10^{7}$ samples. We discarded the first 0.01 $\%$ of MCMC simulations for the ``burn-in'' process. Priors used in the fitting are represented in Table \ref{tab:priors}. In order to verify the dependency of solutions onto assumed priors, we have also adopted uniform priors for all the orbital parameters, yielding the results that are very similar to those with our fiducial priors given in Table \ref{tab:priors}. The longest integrated autocorrelation times calculated in emcee provide insights into the convergence of our simulations \citep{foremanmackey13}. The longest integrated autocorrelation times of all parameters in Table \ref{tab:priors} except for $\gamma$ velocity were calculated to be between 11.58 and 13.16, in which the shortest correlation was 11.57 for the mass of the primary star.

The orbital eccentricity of the binary is very large, 0.62, which may be due to the GJ1108B system with Kozai-Lidov mechanism \citep{kozai62, lidov62}.
The timescale of the oscillation can be written as, 
\begin{eqnarray}
  P_{\rm Kozai-Lidov} = P_{\rm inner} \frac{M_{\rm primary}}{M_{\rm outer}} \left[ \frac{a_{\rm outer}}{a_{\rm inner}} \right]^{3} (1-e_{\rm outer}^2)^{3/2}
  \label{eq:kozai-lidov}
\end{eqnarray}
where $P_{\rm X}, \ M_{\rm X}$, and $a_{\rm X}$ respectively indicate orbital period, mass, and semimajor axis with subscripts, ``inner'' for the inner companion and ``outer'' for the outer companion as a perturber \citep{holman97}. In case of the GJ1108A system with $P_{\rm inner}=8\ {\rm year}, \ M_{\rm primary}=0.7\ M_{\odot}$, and $a_{\rm inner}=4\ {\rm AU}$ and brief assumptions for GJ1108B system: $M_{\rm outer}=0.5\ M_{\odot}$ and $a_{\rm outer}=300\ {\rm AU}$, the timescale becomes $4.7(1-e_{\rm outer}^2)^{3/2}$ Myr. Even if GJ1108B has a circular orbit about the primary, the timescale of Kozai-Lidov mechanism is shorter than the age of the GJ1108 hierarchical system, and hence GJ1108Ab may have experienced the secular perturbation and enhanced its eccentricity.


\section{Discussions}
\indent

\subsection{Revisit of the kinematic age for the GJ1108A system} \label{subsec:age_gj1108A}

Due to orbital motions of the system, their true galactic motions were poorly constrained. The $\gamma$ velocity of GJ1108Aa is estimated as a few km larger than its intrinsic value. We employed an offset velocity in the RV curve as true galactic motion along the line of sight, the $\gamma$ velocity of Table \ref{tab:orbparam}. The proper motions in RA and Dec direction were determined by $Gaia$ satellite \citep{gaiadr2}. The $Gaia$ data release 2 is based on the data collected between mid 25 July 2014 (10:30 UTC) and 23 May 2016 (11:35 UTC). Using ``Observation Forecast Tool of $Gaia$\footnote{https://gaia.esac.esa.int/gost/index.jsp}'', we expected observational epochs for GJ1108A. Based on the observational epochs and the orbital solution in Table \ref{tab:orbparam}, the primary orbits their common center of mass with 118 degree in position angle. We approximated the GJ1108A's velocities along the RA and Dec directions by fitting linear functions with the orbital motions during the $Gaia$'s observation duration, finding that $Gaia$'s proper motion has the contamination of 36.2$\pm$6.0 and -0.32$\pm$0.02 mas/yr along the RA and Dec, respectively. Then, the uncertainty is an average deviancy between the expected orbital motion and the fitted linear function. We subtracted the contamination from the $Gaia$'s proper motions for GJ1108A, determining the true proper motions of GJ1108A system are -49.1$\pm$6.1 and -191.8$\pm$0.3 mas/yr in RA and Dec, respectively. The corrected proper motions of GJ1108A are more consistent with those of GJ1108B, -48.72$\pm$0.16 and -208.85$\pm$0.13 mas/yr in RA and Dec, respectively. Using the BANYAN $\Sigma$ calculator\footnote{http://www.exoplanetes.umontreal.ca/banyan/banyansigma.php} \citep{gagne18} with the updated proper motions and a $\gamma$ velocity of GJ1108A (10.1$\pm$0.2 km s$^{-1}$), we obtained the membership probabilities of GJ1108A to young groups, which are 45.6 and 54.4 $\%$ for the Columba and field region, respectively.

\subsection{Dynamical mass and evolutionary models for the GJ1108A system}

The combination of astrometry and RV revealed the dynamical mass of each component, $0.72_{-0.04}^{+0.04} M_{\odot}$ and $0.30_{-0.03}^{+0.03} M_{\odot}$ for the primary and the companion, respectively. Whereas the recent evolutionary model \citep[][hereafter called BHAC15]{baraffe15} suggests the model-derived masses as $0.68\pm0.01 M_{\odot}$ and $0.23\pm0.02 M_{\odot}$ based on their NIR flux and the age of Columba, in which the mass of both components are under-predicted and the older age may be preferred for GJ1108A system (Figure \ref{fig:masscomp_gj1108A}). The components in the system are here considered as coeval in the mass derivation using the evolutionary model. The right panel of Figure \ref{fig:masscomp_gj1108A} shows model-derived age distributions of the primary and the companion. The primary may be nearly on main-sequence, and the age is not determined well.

We have obtained the properties of GJ1108Aa: flux, dynamical mass, and effective temperature ($T_{\rm eff}=4100_{-400}^{+200}$, See Appendix \ref{sec:app}). Using effective temperature and $H$-band flux, as similar to age estimation via mass and flux based on the evolutionary model, we estimated the isochronal age of 28$\pm$20 Myr, which is inconsistent with the model-derived age of GJ1108Ab, 69$\pm$15 Myr (Figure \ref{fig:masscomp_gj1108A}). The model may predict a higher $T_{\rm eff}$ of the primary or lower luminosity of the companion. The over-prediction of $T_{\rm eff}$ has also recognized on eclipsing binaries \citep[e.g., Figure 11 of ][]{irwin11}.

If the orbital inclination of GJ1108Ab is assumed for the stellar inclination of GJ1108Aa, combining with the rotational period, a stellar radius of the primary can be estimated to be, $v_{\rm rotation} \times P_{\rm rotation}$ $\sim$ 1.2 $R_{\odot}$ ($v_{\rm rotation}$ = $v$sin$i_{\rm star}$ / sin$i_{\rm orbit}$). This estimate is inconsistent with model prediction, which is 0.7 $R_{\odot}$ for a star with the mass of 0.7 $M_{\odot}$ at 40 Myr. Even if GJ1108Aa is at 10 Myr, BHAC15 predicts $\sim$1.0 $R_{\odot}$ for the radius of the primary, which may suggest that the orbital inclination is misaligned with respect to the stellar inclination of the primary due to a perturbation such as Kozai-Lidov mechanism.

Consequently, the system is indeed young as following properties: short rotational period, high UV and X-ray luminosity, kinematics, and the mass-luminosity relation determined by the orbital characterization, indicating the system is younger than 220 Myr. The little lithium abundance of GJ1108Aa should put a constraint on lower limit for their age (20Myr). In summary with current knowledge, a membership to Columba or GJ1108A is a young system in a field region are probable. Then, we here tentatively employ the age of Columba moving group, $\ageCOL$ Myr for the GJ1108A system as the age less independently determined to evolutionary models.


\begin{deluxetable*}{ccc ccc ccc}
\centering
\tablewidth{0pt}
\tablecaption{Physical parameters for the comparison of the dynamical mass and the model-derived mass}
\tablehead{
    \colhead{Name} &
    \colhead{Reference\tablenotemark{a}} &
    \colhead{distance} &
    \colhead{age} &
    \colhead{mass$_{\rm total}$} &
    \colhead{mass$_{\rm prim}$} &
    \colhead{mass$_{\rm comp}$} &
    \colhead{separation} &
    \colhead{tertiary} \\
    \colhead{} & \colhead{} & \colhead{[pc]} & \colhead{[Myr]} & 
    \colhead{[$M_{\odot}$]} & \colhead{[$M_{\odot}$]} & \colhead{[$M_{\odot}$]} & \colhead{[AU]} & \colhead{}
}
\startdata
AB DorA ab     & C05 & 15.18$\pm$0.13 & $\ageABD$\tablenotemark{b} & - & 0.865$\pm$0.034 & 0.090$\pm$0.005 & 3.07$\pm$0.15 & AB DorB \\
AB DorB ab     & A15 & 15.18$\pm$0.13 & $\ageABD$\tablenotemark{b} & - & 0.28$\pm$0.05   & 0.25$\pm$0.05   & 0.79$\pm$0.03 & AB DorA \\
TWA 5A a+b     & K07 & 48.7$\pm$5.7\tablenotemark{c1} & $\ageTWH$\tablenotemark{b} & 0.96$\pm$0.19   & - & - & 3.21$\pm$0.45 & TWA 5B \\
TWA 22 a+b     & B09 & 17.5$\pm$0.2\tablenotemark{c2} & $\ageBPC$\tablenotemark{b, c2} & 0.220$\pm$0.021 & - & - & 1.77$\pm$0.04 & - \\
HD 130948 b+c  & D09 & 18.17$\pm$0.11 & 790$_{-150}^{+220}$      & 0.109$\pm$0.003 & -                     & - & 2.20$\pm$0.11 & HD 130948 \\
HR 7672 b      & C12 & 17.77$\pm$0.11 & 2400$_{-700}^{+600}$     & -               & 1.08$\pm$0.04\tablenotemark{d} & 0.069$_{-0.003}^{+0.002}$ & 18.3$_{-0.5}^{+0.9}$ & - \\     
Gl 417 b+c     & D14 & 21.93$\pm$0.21 & 750$_{-120}^{+150}$      & 0.099$\pm$0.003 & -                     & - & 2.85$\pm$0.03 & Gl 417 \\                                
GJ 3305 ab     & M15 & 29.43$\pm$0.30 & $\ageBPC$\tablenotemark{b} & 1.10$\pm$0.04 & 0.65$\pm$0.05 & 0.44$\pm$0.05   & 9.78$\pm$0.14 & 51 Eri \\
V343 Normae ab & N16 & 38.54$\pm$1.69 & $\ageBPC$\tablenotemark{b} & 1.39$\pm$0.11 & 1.10$\pm$0.10 & 0.290$\pm$0.018 & 3.07$\pm$0.08 & HD 139084B \\  
HD 4747 b      & C16 & 18.69$\pm$0.19 & 3300$_{-1900}^{+2300}$   & -               & 0.82$\pm$0.04\tablenotemark{d} & 0.060$\pm$0.003 & 16.4$_{-3.3}^{+3.9}$ & - \\
GJ 1108A ab    & -   & 24.83$_{-0.22}^{+0.22}$ & $\ageCOL$\tablenotemark{b}        & - & $0.72_{ -0.04}^{+ 0.04}$ & $0.30_{ -0.03}^{+ 0.03}$ & $4.11_{ -0.05}^{+ 0.05}$ & GJ 1108B
\enddata
\label{tab:lite}
\end{deluxetable*}

\begin{deluxetable*}{ccc ccc c}
\addtocounter{table}{-1}
\tablecaption{Continued}
\centering
\tablewidth{0pt}
\tablehead{ \colhead{} & \colhead{} & \colhead{} & \colhead{} & \colhead{} & \colhead{} & \colhead{} }
\startdata 
Name           & m$_{J, {\rm prim}}$ & m$_{H, {\rm prim}}$ & m$_{K, {\rm prim}}$ & m$_{J, {\rm comp}}$ & m$_{H, {\rm comp}}$ & m$_{K, {\rm comp}}$ \\ \\
\tableline
AB DorA ab     & 5.32$\pm$0.02\tablenotemark{e} & 4.85$\pm$0.03\tablenotemark{e} & 4.69$\pm$0.02\tablenotemark{e} & 10.76$_{-0.24}^{+0.19}$ & 10.04$_{-0.15}^{+0.13}$ & 9.45$_{-0.15}^{+0.12}$ \\
AB DorB ab     & 8.17$\pm$0.02\tablenotemark{e} & 8.29$\pm$0.04\tablenotemark{f1} & 7.97$\pm$0.03\tablenotemark{f1}  & - & 8.55$\pm$0.04\tablenotemark{f1} & 8.23$\pm$0.03\tablenotemark{f1} \\    
TWA 5A a+b     & 8.40$\pm$0.07    & 7.69$\pm$0.04   & 7.39$\pm$0.04   & 8.45$\pm$0.15  & 7.79$\pm$0.05  & 7.62$\pm$0.08 \\
TWA 22 a+b     & 9.12$\pm$0.10    & 8.61$\pm$0.15   & 8.24$\pm$0.19   & 9.52$\pm$0.11  & 9.12$\pm$0.15  & 8.70$\pm$0.25 \\
HD 130948 b+c  & 13.81$\pm$0.06   & 13.04$\pm$0.10  & 12.26$\pm$0.03  & 14.12$\pm$0.06 & 13.33$\pm$0.11 & 12.46$\pm$0.03 \\      
HR 7672 b      & 4.69\tablenotemark{e} & 4.43\tablenotemark{e} & 4.39$\pm$0.03\tablenotemark{e} & 14.39$\pm$0.20\tablenotemark{f2} & 14.04$\pm$0.14\tablenotemark{f2} & 13.04$\pm$0.10\tablenotemark{f2} \\
Gl 417 b+c     & 15.05$\pm$0.04   & 14.19$\pm$0.05  & 13.29$\pm$0.03  & 15.49$\pm$0.04 & 14.45$\pm$0.06 & 13.63$\pm$0.03 \\     
GJ 3305 ab     & 7.67$\pm$0.02    & 7.01$\pm$0.05   & 6.80$\pm$0.02   & 8.64$\pm$0.02  & 8.00$\pm$0.05  & 7.73$\pm$0.02 \\           
V343 Normae ab & 6.44$\pm$0.12    & 6.05$\pm$0.10   & 5.93$\pm$0.11   & 9.69$\pm$0.12  & 9.20$\pm$0.10  & 8.79$\pm$0.11 \\
HD 4747 b      & 5.81$\pm$0.02\tablenotemark{e} & 5.43$\pm$0.05\tablenotemark{e} & 5.31$\pm$0.03\tablenotemark{e} & - & - & 14.36$\pm$0.14 \\     
GJ 1108A ab    & 7.37$\pm$0.02    & 6.74$\pm$0.02   & 6.55$\pm$0.03   & 9.34$\pm$0.05  & 8.74$\pm$0.04  & 8.55$\pm$0.03
  \enddata
\tablenotetext{a}{Reference papers to determine the dynamical mass are shown, in which many of the parameters including distance, age, and the luminosities are presented unless otherwise noted. Abbreviations indicate: C05 for \cite{close05}, A15 for \cite{azulay15}, K07 for \cite{konopacky07a}, B09 for \cite{bonnefoy09}, D09 for \cite{dupuy09b}, C12 for \cite{crepp12b}, D14 for \cite{dupuy14}, M15 for \cite{montet15}, C16 for \cite{crepp16}, and N16 for \cite{nielsen16}.}
\tablenotetext{b}{The age of a moving group is employed, $\ageABD$ Myr for AB Doradus, $\ageTWH$ Myr for TW Hydrae, $\ageBPC$ Myr for $\beta$ Pictoris, and $\ageCOL$ Myr for Columba \citep{bell15}.}
\tablenotetext{c1}{The distance of stars in the TW Hydrae association is referred from \cite{ducourant14}.}
\tablenotetext{c2}{The distance and membership of TWA 22 are referred from \cite{teixeira09}.}
\tablenotetext{d}{Non-dynamical mass determined by empirical tracks, or theoretically determined using spectroscopic results using tools such as SME \citep{valentipiskunov96, valentifischer05}.}
\tablenotetext{e}{Unresolved 2MASS photometry for those systems \citep{skrutskie06}.}
\tablenotetext{f1}{Photometric measurements are obtained from \cite{janson07a}.}
\tablenotetext{f2}{Photometric measurements are obtained from \cite{boccaletti03}.}
\end{deluxetable*}


\begin{figure*}
  \centering
  \begin{tabular}{ccc}
     \begin{minipage}[htbp]{0.33\hsize}
       \includegraphics[width=\hsize]{\dirGJ1108A/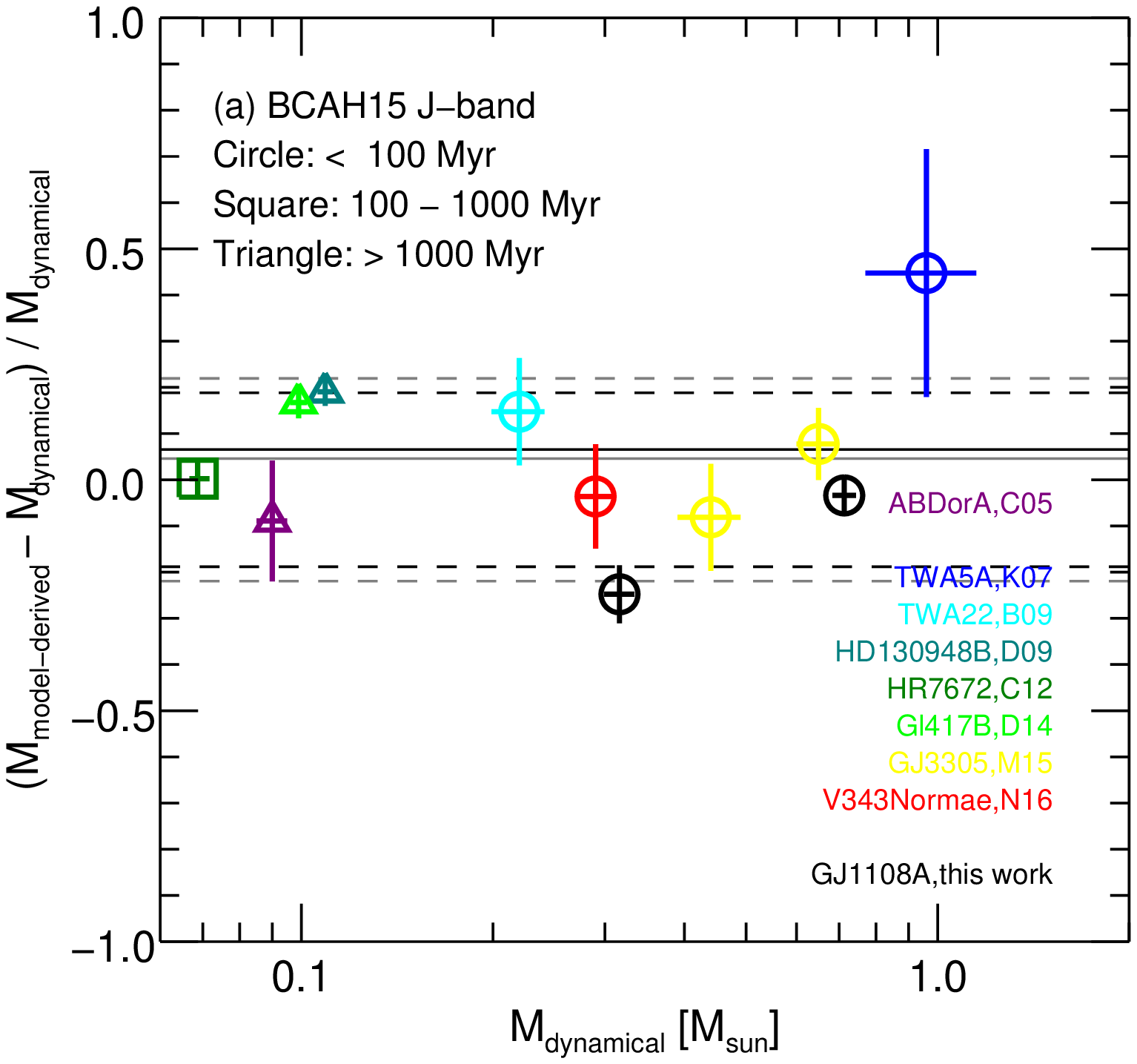}
     \end{minipage}
     \begin{minipage}[htbp]{0.33\hsize}
       \includegraphics[width=\hsize]{\dirGJ1108A/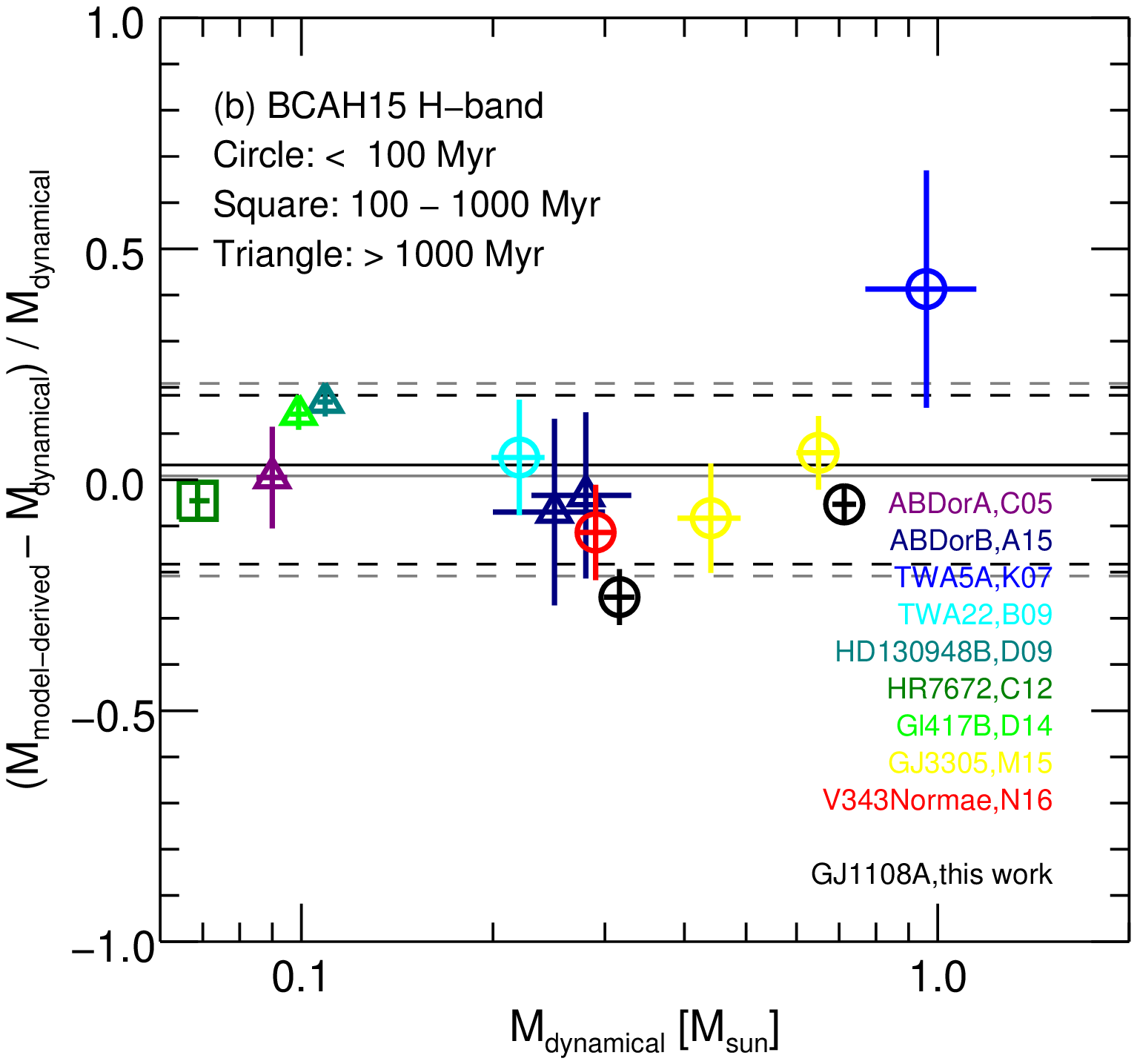}
     \end{minipage} 
     \begin{minipage}[htbp]{0.33\hsize}
       \includegraphics[width=\hsize]{\dirGJ1108A/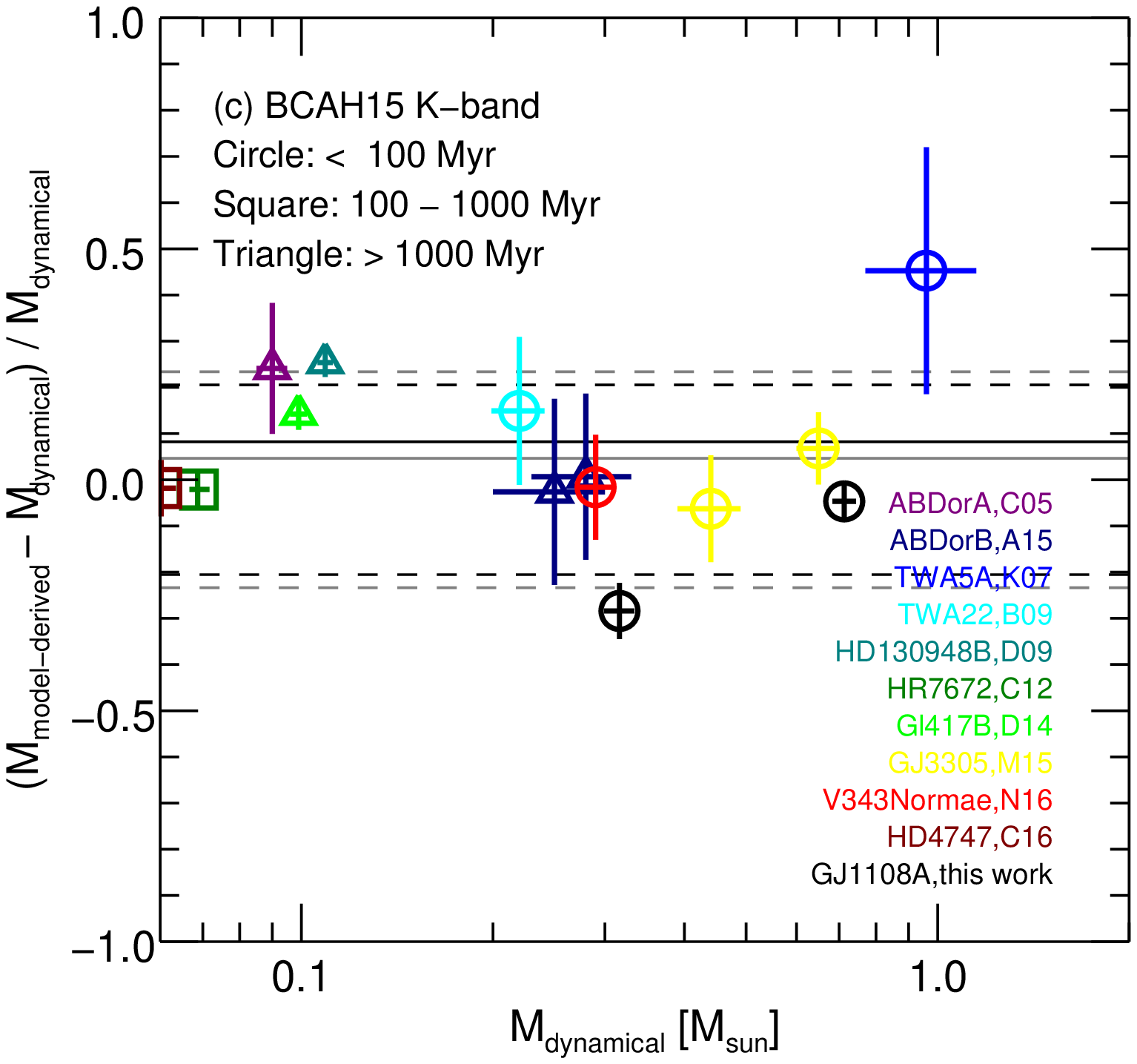}
     \end{minipage}
  \end{tabular}
  \caption{The dynamical mass and model-derived mass based on BHAC15 model are compared on the figures. Data points are obtained from \cite{close05, konopacky07a, bonnefoy09, dupuy09b, crepp12b, dupuy14, azulay15, montet15, crepp16, nielsen16}, and classified according to their age as symbols with different colors. To estimate the model-derived mass, $J$, $H$, and $K$-band photometry are used in each panel (a), (b), and (c), respectively. The black solid and dashed lines on those show a mean offset and scatter, and the grey lines are for only resolved PMS stars.}
  \label{fig:masscomp_b15}
\end{figure*}

\subsection{What causes the mass discrepancy ?}
\subsubsection{Accretion history}
Several studies have been indicated the importance of accretion histories before the quasi-static contraction phase \citep[e.g.][]{baraffe09, baraffe12}. Due to the different types of accretion mechanisms, conditions of young low-mass stars can become much different from those considered in classical models. One of the most important subjects regarding accretion histories is the balance between expansion and contraction for a protostar, which strongly depends on details of accretion processes. However the details are still remained very uncertain. To quantitatively investigate the accretion effect: the energy loss during the accretion onto protostar's surface, a free parameter $\alpha$ have been employed in \cite{hartmann97}. The increase of internal energy for protostar and the radiation at its surface are respectively written as,
\begin{eqnarray}
E_{\rm add} & = & \alpha \epsilon \frac{GM}{R} \dot{M} \\
E_{\rm rad} & = & (1-\alpha) \epsilon \frac{GM}{R} \dot{M}
\end{eqnarray}
where $G, \ M, \ R$, and $\dot{M}$ indicate gravitational constant, mass of protostar, radius of protostar, and mass of accreted matter. The $\epsilon$ presents the fraction of internal energy retained in an accretion disk, up to 0.5. In case of $E_{\rm add}$ is dominant, a protostar will expand due to the accreted energy and become brighter than non-accreted objects in the models, and hence such accretion is called ``hot accretion''. In the opposite case, a protostar becomes smaller and fainter due to ``cold accretion''. The threshold of $\alpha$ is obtained by the energy equation \citep[Eq.6 of ][]{hartmann97}, and $\alpha$ of 0.1--0.2 have been assumed \citep{hartmann97, baraffe09}.

If we assume that GJ 1108A is a member of the Columba group, cold accretion is preferred in its accretion history to explain the mass under-prediction. we assume that GJ1108A is a member of the Columba group, cold accretion is preferred in its accretion history to explain the mass under-prediction. However young stars may forget the accretion history at the age of GJ1108A, 40--50 Myr \cite[Figure 1 of ][]{baraffe10}, and the reason why such significant events had occurred in the system is a matter to be further investigated.

\subsubsection{Magnetic activity}
Magnetic activities are also essential to understand the stellar structure and their observable properties. Similar discrepancy to evolutionary models has been recognized on eclipsing binaries in mass-radius relation \citep[e.g.][]{irwin11}, and theoretical investigations suggest that models considering magnetic activities may explain the discrepancy \citep{feiden13a}. The magnetic field works as inhibition of stellar convection, making spots at stellar surface and making stellar radius inflated compared with classical models. In other words, active low-mass stars tend to be observed to classical models as larger radius for given $T_{\rm eff}$ or lower $T_{\rm eff}$ for given radius \citep[e.g.][]{mullanmacdonald01}. \cite{feiden13b, feiden14} demonstrate that stellar evolutionary code can explain the discrepancy by involving magnetic activity for both stars with a radiative core and fully convective stars. This approach also better reproduces the observed color-magnitude diagram of young stellar association with single isochrone \citep{feiden16}.

GJ1108Aa is a rapid rotating and indeed X-ray luminous young system. If we assume the membership to Columba for GJ1108A, BHAC15 model not including the magnetic activity predicts its flux brighter than observed flux. The observed $T_{\rm eff}=4100_{-400}^{+200}$ is marginally consistent with the model prediction (4084 K) for 0.7 $M_{\odot}$ at 40 Myr. To explain the observed deviation, the primary must be smaller than the classical model prediction, which is opposite to the behavior of magnetic activity, ``a larger radius for a given $T_{\rm eff}$''. Hence, a scenario that GJ1108A is older than the age of Columba is more preferred than the non-included effects to classical models such as magnetic inhibition of convection or accretion history (See the previous section).

\begin{deluxetable*}{ccc ccc cc}
\centering
\tablewidth{0pt}
\tablecaption{Results of the mass comparison}
\tablehead{ \colhead{Model set} & \colhead{Mean offset\tablenotemark{a}} & \colhead{Scatter\tablenotemark{a}} &
  \colhead{Method} & \colhead{Mass range} & $N_{\rm sample}$ & Tertiary rate & \colhead{Reference} \\
  \colhead{} & \colhead{} & \colhead{} & 
  \colhead{} & \colhead{$M_{\odot}$} & \colhead{} & \colhead{} & \colhead{} }
\startdata
BHAC15 (All,$J$)                  &  0.07$\pm$0.04 & 0.19$\pm$0.05 & age and $L_{J}$ & 0.05--0.7 & 11 & 6/9 & this work \\
BHAC15 (All,$H$)                  &  0.03$\pm$0.04 & 0.18$\pm$0.05 & age and $L_{H}$ & 0.05--0.7 & 12 & 6/9 & this work \\ 
BHAC15 (All,$K$)                  &  0.08$\pm$0.04 & 0.21$\pm$0.04 & age and $L_{K}$ & 0.05--0.7 & 13 & 7/10& this work \\

BHAC15 (PMS,$J$)                  & -0.05$\pm$0.04 & 0.17$\pm$0.04 & age and $L_{J}$ & 0.09--0.65 & 6 & 4/4 & this work \\
BHAC15 (PMS,$H$)                  & -0.05$\pm$0.04 & 0.18$\pm$0.04 & age and $L_{H}$ & 0.09--0.65 & 8 & 5/5 & this work \\ 
BHAC15 (PMS.$K$)                  &  0.08$\pm$0.04 & 0.21$\pm$0.04 & age and $L_{K}$ & 0.09--0.65 & 8 & 5/5 & this work \\

BCAH98\tablenotemark{b} (All,$J$) &  0.01$\pm$0.06 & 0.17$\pm$0.05 \\
BCAH98\tablenotemark{b} (All,$H$) &  0.02$\pm$0.06 & 0.17$\pm$0.04 \\
BCAH98\tablenotemark{b} (All,$K$) &  0.03$\pm$0.04 & 0.21$\pm$0.03 \\
\tableline
BCAH98\tablenotemark{b}           & 0.342$\pm$0.471 & 0.439           & $T_{\rm eff}$ and $L$ & 0.03--1.4 & 16 & 5/8 & S14\tablenotemark{d} \\
BCAH98\tablenotemark{c}           & 0.212$\pm$0.353 & 0.308           &
\enddata
\tablenotetext{a}{The mean offset and scatter in the mass comparison are determined by a typical value of $(M_{\rm model}-M_{\rm dynamical})/M_{\rm dynamical}$.}
\tablenotetext{b}{[M/H]=0, $l_{\rm mix}=1.0H_{p}$, and Y=0.275}
\tablenotetext{c}{[M/H]=0, $l_{\rm mix}=1.9H_{p}$, and Y=0.282}
\tablenotetext{d}{\cite{stassun14}}
\label{tab:offset}
\end{deluxetable*}

\begin{figure*}
  \centering
  \includegraphics[width=\hsize]{\dirGJ1108A/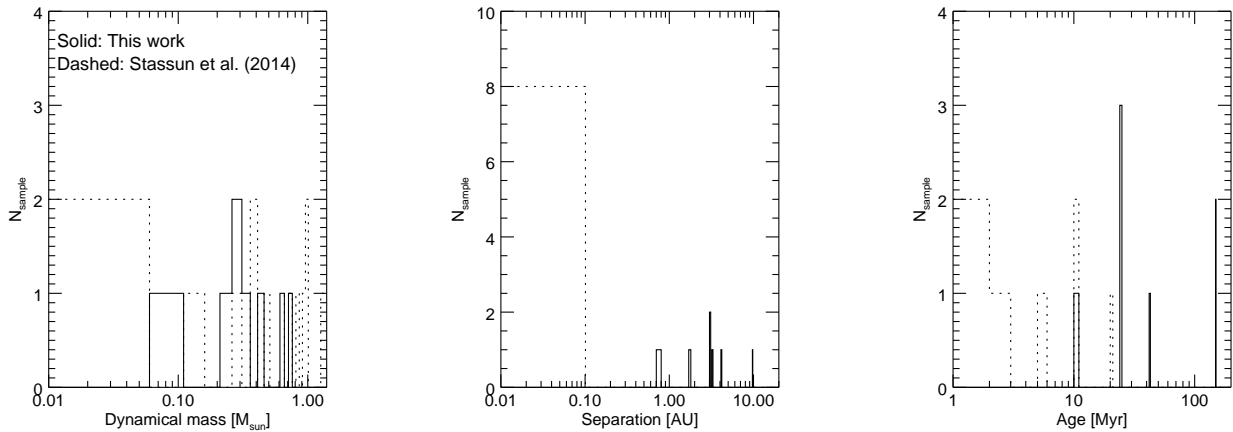}
  \caption{The number of object focused on properties: mass, separation, and age of systems are shown. Solid and dashed histograms indicate resolved PMS stars compiled in this work and all the low-mass systems of \cite{stassun14}, respectively.}
  \label{fig:samples}
\end{figure*}

\subsection{Mass comparison using compiled samples}
\subsubsection{Sample selection}
In order to verify the performance of present evolutionary models, especially the luminosity evolution observed in NIR wavelength as a function of time, we compare the dynamical masses of stars with the model-derived masses of those (Figure \ref{fig:masscomp_b15}), by collecting information from the literature. Physical parameters of the literature-based samples such as age, dynamical mass, and broad-band fluxes, are summarized in Table \ref{tab:lite}.

We set three criteria for samples in the mass comparison. As the first criterion, we selected objects whose mass are dynamically estimated. The mass of each component has not separately measured for some of our sample; their orbital RV modulations have not been well determined. We then simply employ the total masses of two components in the systems, which were determined by astrometry, in our mass comparisons.

Second, we select the objects that have been age-dated. At this step, we do not adopt age dating based on isochrone for a single stellar system. If a sample is associated with any stellar cluster or moving group, the sample' age can be represented by a well-determined isochronal age of the coeval population in groups; our sample therefore includes the objects that are associated to stellar clusters or moving groups.

Third, we here focus on PMS stars, since the tests of evolutionary models for the PMS phase should be more incomplete than the MS phase. Then, we do not include the members of star-forming regions, for which the stellar properties such as distance, brightness, and age are relatively poorly determined compared with post T-tauri stars in the nearby stellar groups.
Age-dated brown dwarfs\footnote{As the age of brown dwarfs in our mass comparison (Table \ref{tab:lite}), we employed the age of stellar primary (tertiary). } are also included in our sample, because this study is motivated for better understanding brown dwarfs and giant planets based on evolutionary models. It should be noted that the mass comparison is conducted for both only PMS stars and all the sample including PMS stars and brown dwarfs.

\subsubsection{Dynamical mass vs Model-derived mass}
To estimate the masses of the samples, we need to select the fiducial sets of evolutionary models. For low-mass objects, non-grey atmosphere is crucial to properly reproduce the stellar structure as already discussed in \cite{chabrierbaraffe97}. We provide the following two criteria; the models (1) cover a wide mass range from brown dwarfs to stars and (2) adopt non-grey atmosphere. \cite{stassun14} reviewed several theoretical isochrones made by different groups, from which we selected the latest model of the Lyon model \citep{baraffe98, baraffe15}.

The dynamical masses and model-derived masses are compared, and we evaluated an offset and scatter between the two masses. We ran a Monte-Carlo simulation in order to derive the uncertainties of the offset and scatter value. This simulation randomly generated masses of all the samples based on their errors, to make a distribution of those values and calculates the typical values from the distribution. Furthermore, the number of resolved young binary is still small, which may make the mass comparison unreliable. To verify this problem, we randomly omit 10--50$\%$ of samples and carried out the same comparison as described above. The offsets and scatters with those partially-selected samples are the almost same as what was derived using all the samples.

The mass comparison of Figure \ref{fig:masscomp_b15} respectively results with small offset and scatter, $\le10\%$ and $\sim20\%$, indicating the luminosity evolution for low-mass objects are well reproduced on average by the evolutionary model of \cite{baraffe15}. Even if we exclude brown dwarfs in the mass comparison, similar results could be obtained. The old version of the model \cite[][hereafter BCAH98]{baraffe98} is also adopted for the mass comparison. Although BCAH98 has three model grids with different mixing length parameter, $l/H_{p}$=1.0, 1.5, and 1.9, the latter two grids do not have the mass range for brown dwarfs, and we used only the grid of $l/H_{p}$=1.0 in this work. We used the same approach as for BHAC15, providing a similar result between these two models. These are summarized in Table \ref{tab:offset}. 

\subsection{Comparison with previous study}
\cite{stassun14} investigated the performance of stellar evolutionary models including BCAH98 using literature-based PMS eclipsing binaries whose isochronal age is spanned in 1--20 Myr. Using luminosity and effective temperature, they estimated the model-derived mass for 16 objects of 0.03--1.4$M_{\odot}$ with the BCAH98 model, and then conducted the mass comparison. They found the model overestimate the dynamical mass by 34.2$\pm$47.1 and 21.2$\pm$35.3$\%$ that are dependent on the mixing length parameter. For the mass discrepancy, they considered the mainly three possibilities: activities, initial events in star formation, and tertiary. Whereas significant improvements were not obtained due to corrections of the magnetic activity and accretion history for evolutionary models, the discrepancy might be seen for systems possessing a tertiary. Tertiaries may occur the non-negligible deviations in the mass comparison as they suggested.
On the other hand, the evolutionary model has better performance for our resolved binaries. Figure \ref{fig:samples} compares properties of eclipsing binaries of \cite{stassun14} and resolved binaries in this work, focusing on mass, separation, and age. The resolved binaries in this work typically are older and have the larger separation, which may attenuate the the discrepancy of mode-derived masses from dynamical masses.


\section{Summary}
\indent

Orbital characterization of GJ1108A system is presented using high-resolution data of the Subaru Telescope combined with archival datasets. The relative orbit of the system is highly eccentric with $e$=0.63. The timescale of Kozai-Lidov mechanism is shorter than the age of GJ1108A system, several tens Myr, and hence the widely separated system, GJ1108B might work as a perturber. If we assume the membership to Columba for GJ1108A system, the BHAC15 evolutionary model estimated masses are slightly larger. Some effects unimplemented in classical models are not preferred to explain the mass discrepancy, because GJ1108A is too old that accretion history accounts for the discrepancy and its luminosity is low compared with the observed $T_{\rm eff}$ in terms of magnetic activity. Hence the GJ1108A system may be older than the Columba moving group, about 69$\pm$15 Myr if the mass-luminosity relation for GJ1108Ab based on evolutionary models is correct, which is still consistent with the age range suggested by spectroscopic and activity age indicators, 20--220 Myr. Combined with the literature-based benchmark stars, the performance of evolutionary models is also discussed. Consequently, the BHAC15 model on average reproduces the mass-luminosity relation of M-dwarfs and massive brown dwarfs at different stages of their evolution. The performance for young resolved binaries is a few times better than that for young eclipsing binaries, which may come from differences in separation and age of binary systems used for the mass comparison.

There are still observational questions, the number of benchmark star and the methods to estimate stellar properties, to understand evolution of low-mass objects and models for them. In order to improve the understanding of contracting and cooling evolution for low-mass objects including exoplanets discovered in near future, orbital characterizations for benchmark stars should be an important step. The $Gaia$'s survey is going to reveal and accurately characterize the low-mass population in the solar vicinity, young binaries of which will be follow-up by ground-based high-resolution instruments for their orbital motions. Increasing the benchmark stars enable to advance the understanding of the young stellar evolution, and extend to the cooling evolution of brown dwarf and giant planet, potentially.
\newline
\newline
Facilities: \facility{Subaru Telescope}, \facility{Keck}, \facility{Observatoire de Haute-Provence}, \facility{Apache Point Observatory}.



\appendix

\section{Grism spectroscopy for the primary of GJ1108A system} \label{sec:app}
\indent

In order to spectroscopically characterize the GJ1108A system, AO observations using Subaru/IRCS were conducted as a part of the SEEDS survey. Frames were taken in ABBA nodding. Spectral resolution of IRCS-grism is determined by the pixel scale and slit width. We employed the pixel scale of 52 mas and Reflective 4 slit with a width of 0.30$\arcsec$, producing $\lambda/\Delta\lambda=$ 477 and 382 in $J$ and $H$-band, respectively. Spectra of Ar lamp was taken at the end of the observation for wavelength calibration. In the observation, only GJ1108Aa was on the slit, and HD68933 was observed for the telluric absorption correction. These observing properties are summarized in Table \ref{tab:grism}.

The initial image processing consisted of flat fielding, subtracting the pairs of images taken at the two different position on a detector, subtracting the sky residual estimated on the background region, and wavelength calibration with Ar emission lines. Flux scaling was performed for each one-dimensional spectra, based on the mean counts in the wavelength coverage. All the spectra were median combined to an image-processed spectrum.

\begin{deluxetable*}{ccc ccc ccc}[htbp]
\centering
\tablewidth{0pt}
\tablecaption{Observing properties of grism spectroscopy with IRCS}
\tablehead{
  \colhead{Obs. Date (UT)} &
  \colhead{Instrument} &
  \colhead{Object} &
  \colhead{$\lambda$} & 
  \colhead{$\lambda/\Delta\lambda$} &
  \colhead{$N_{\rm exp}(N_{\rm coadd})$} &
  \colhead{$t_{\rm tot}$} &
  \colhead{Airmass} &
  \colhead{FWHM} \\
  \colhead{yyyy-dd-mm} & \colhead{} & \colhead{} & \colhead{} & \colhead{} &
  \colhead{} & \colhead{[s]} & \colhead{} & \colhead{[mas]}
}
\startdata
2012-12-04 & IRCS/grism & GJ1108Aa & $J$ & 477 & 8(10) & 120 & 1.07 & 128.37 \\
           &            &          & $H$ & 382 & 8(10) & 120 & 1.06 & 125.29 \\
           &            & HD68933  & $J$ & 477 & 4(30) &  84 & 1.08 & 152.75 \\
           &            &          & $H$ & 382 & 4(30) &  49 & 1.07 & 151.32
\enddata
\label{tab:grism}
\end{deluxetable*}

For the correction of telluric absorption, HD68933 identified as an F5 dwarf on the SIMBAD database \citep{wenger00} was used. The effective temperature of the star is summarized here based on the literature. \cite{masana06} estimated it as 6123$\pm$58 K with $V$-band and 2MASS infrared photometries. On the other hand, \cite{casagrande11} and \cite{mcdonald12} indicate their $T_{\rm eff}$ as 6209 K by the Genevacopehagen survey and 6000 K by multi-band photometry, respectively. We here employ $T_{\rm eff}$ of the telluric standard star as 6100$\pm$100 K. It may be difficult to separate intrinsic metal lines of the standard star from telluric absorption lines in such low-resolution spectra. To minimize the imperfection of telluric correction, we compared observed spectra with a model spectrum as a true spectrum of the standard, BT-settle model \citep{allard12}, convolved to observing resolution. 

\begin{figure*}[htbp]
  \centering
  \begin{tabular}{cc}
     \begin{minipage}[t]{0.45\hsize}
       \includegraphics[width=\hsize]{\dirGJ1108A/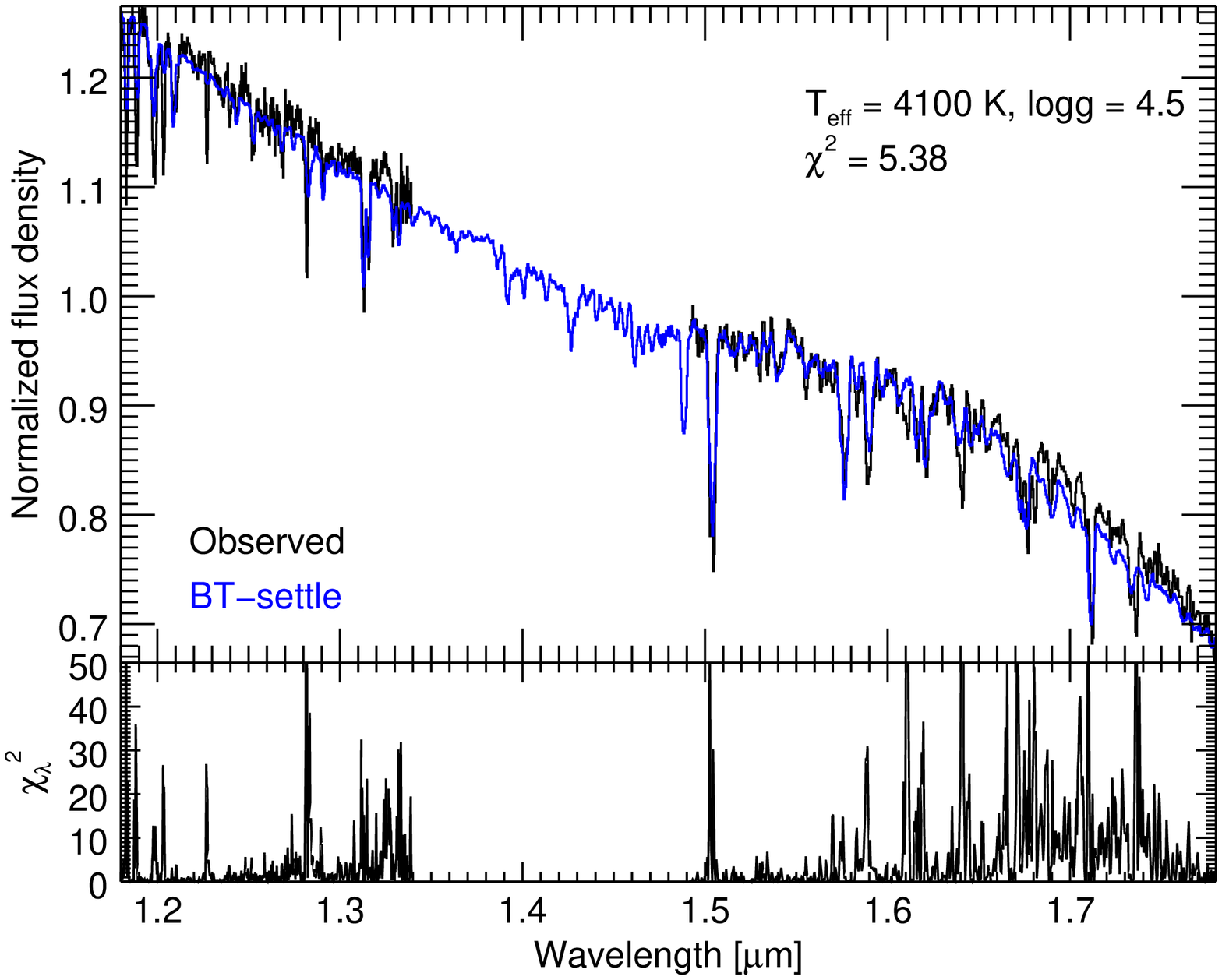}
     \end{minipage}
     \begin{minipage}[t]{0.45\hsize}
       \includegraphics[width=\hsize]{\dirGJ1108A/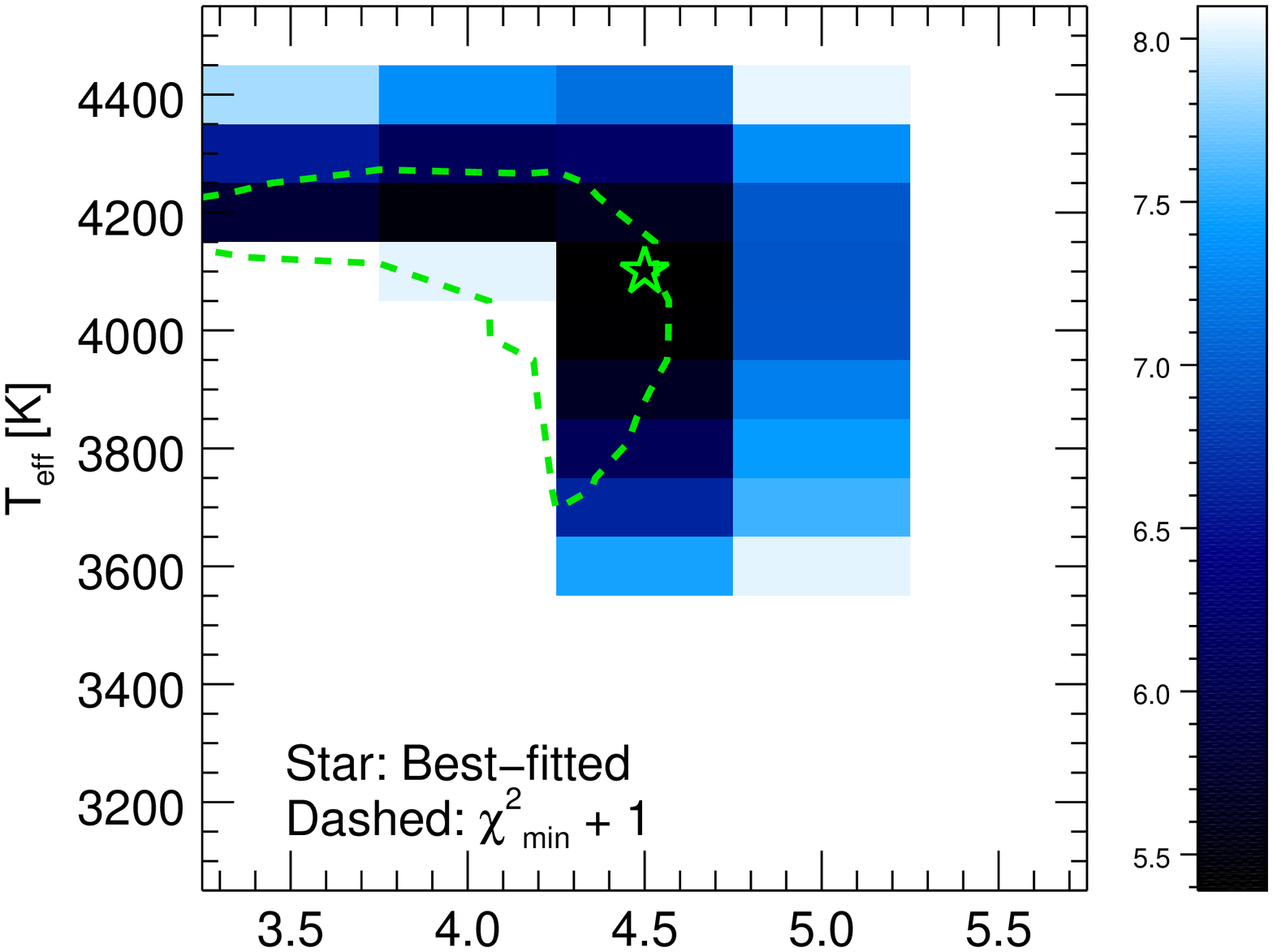}
     \end{minipage}
   \end{tabular}
  \caption{The left presents an observed $JH$-band spectra and a best-fitted model, in which chi-square of each wavelength are also shown. The right shows the reduced chi-square map of the fitting.}
  \label{fig:jhspectra}
\end{figure*}

The reduced spectrum were fitted with BT-settle models on reduced chi-square map of $T_{\rm eff}$ and log $g$, indicating $T_{\rm eff}=4100_{-400}^{+200}$ K with log $g = 4.5_{-1.0}^{+0.0}$ dex for the effective temperature and surface gravity of GJ1108A (Figure \ref{fig:jhspectra}). Even if we use each broad-band spectrum ($J$- or $H-$band of IRCS-grism) for the fitting, similar results were obtained. Since the companion GJ1108Ab is very close to the primary and their contrast is also small, the spectra may be contaminated due to the companion. In addition, the observing condition such as AO performance might be variable. Follow-up observations are strongly required for detailed characterizations of the system.

As a comparison, we estimated the color temperature of GJ1108Aa with optical--infrared photometries. The optical flux of the primary must be contaminated by the companion. Using BHAC15 evolutionary model with inputs of $0.30\pm0.03 M_{\odot}$ and 20--220 Myr, model-dependent $V-JHK$ colors and $V$-band contrast of about a factor of 15$\pm$1 times are obtained. The color temperature of GJ1108Aa based on \cite{casagrande08} is $T_{\rm color}=3897\pm110$ K, which is roughly consistent with the spectroscopic temperature of $T_{\rm eff}=4100_{-400}^{+200}$. The effective temperature of $Gaia$ DR2 ($T_{\rm eff}=3989.2_{-422.9}^{+218.2}$) is also marginally consistent with our result.

\bibliographystyle{apj}  
\bibliography{mizuki17} 

\acknowledgments
The authors thank the anonymous referee for useful comments and the Subaru Telescope staff for their assistance.
This work was conducted based on: 
(a) data collected at Subaru Telescope, which is operated by the National Astronomical Observatory of Japan, especially ``Strategic Exploration of Exoplanets and Disks with Subaru'';
(b) the data taken by SOPHIE of Observatoire de Haute-Provence 1.93 meter telescope and archived in their library;
(c) observations obtained with the Apache Point Observatory 3.5-meter telescope, which is owned and operated by the Astrophysical Research Consortium;
(d) the data obtained at the W.M. Keck Observatory, which is operated as a scientific partnership among the California Institute of Technology, the University of California and the National Aeronautics and Space Administration. The Keck observatory was made possible by the generous financial support of the W.M. Keck Foundation.
The authors wish to recognize and acknowledge the very significant cultural role and reverence that the summit of Mauna Kea has always had within the indigenous Hawaiian community. We are most fortunate to have the opportunity to conduct observations from this mountain.
This research has made use of the SIMBAD database, operated at CDS, Strasbourg, France.
The distortion correction applied to the HiCIAO data benefited from observations made with the NASA/ESA Hubble Space Telescope, and obtained from the Hubble Legacy Archive, which is a collaboration between the Space Telescope Science Institute, the Space Telescope European Coordinating Facility (ST-ECF/ESA) and the Canadian Astronomy Data Centre (CADC/NRC/CSA).
Data analysis were in part carried out on common use data analysis computer system at the Astronomy Data Center, ADC, of the National Astronomical Observatory of Japan.
T.M. was supported by the Program for Leading Graduate Schools, ``Inter-Graduate School Doctoral Degree Program on Global Safety'', by the Ministry of Education, Culture, Sports, Science, and Technology.
This work was partially supported by the Grant-in-Aid for JSPS Fellows (Grant Number 25-8826).
This work was supported by Japan Society for Promotion of Science (JSPS) KAKENHI Grant Number JP16K17660.
This research was partly supported by JSPS KAKENHI Grant Number JP18H01265.
This work was also supported by MEXT KAKENHI No. 17K05399 (E.A.).
\end{document}